\documentclass[10pt, twocolumn]{IEEEtran}
\usepackage{graphicx,amsmath,amssymb,cite}
\usepackage{subfigure}
\usepackage{mathabx}
\usepackage{float}
\usepackage{graphicx,subfigure,amsmath,amssymb,amsfonts,cite,float,color,longtable,rotating,diagbox,booktabs,array}
\makeatletter

\newcommand{\Rmnum}[1]{\expandafter\@slowromancap\romannumeral #1@}
\makeatother
\usepackage{float,color}
\ifCLASSINFOpdf
\else
\fi

\begin{document}

\title{Machine-Learning-based High-resolution DOA Measurement and Robust DM for Hybrid  Analog-Digital Massive MIMO Transceiver}

\author{Feng~Shu,~\IEEEmembership{Memeber,~IEEE},~Ling~Xu,~Jinyong~Lin,~Jingsong Hu,\\~Linqing Gui,~Jun Li,~and~Jiangzhou~Wang,~\IEEEmembership{Fellow,~IEEE}
\thanks{This work was supported in part by the National Natural Science Foundation of China (Nos. 61472190, and 61501238).}
\thanks{Feng~Shu,~Ling~Xu,~Jinsong~Hu,~Yijin Zhang,~Linqing~Gui,~and~Jun~Li are with School of Electronic and Optical Engineering, Nanjing University of Science and Technology, 210094, CHINA. E-mail:\{shufeng, xuling\}@njust.edu.cn}
\thanks{Feng~Shu,and~Jinsong Hu  are the College of Computer and Information Sciences, Fujian Agriculture and Forestry University, Fuzhou 350002, China (e-mail: shufeng0101@163.com).}
\thanks{Jinyong Lin is with Beijing Aerospace Automatic Control Institute, Beijing 100854, China (e-mail: ljiny3771@sina.com)}
\thanks{Jiangzhou Wang is with the School of Engineering and Digital Arts, University of Kent, Canterbury CT2 7NT, U.K. Email: \{j.z.wang\}@kent.ac.uk.
}
}

\maketitle
\begin{abstract}
At hybrid analog-digital (HAD) transceiver, an improved HAD rotational invariance techniques (ESPRIT), called I-HAD-ESPRIT, is proposed to measure the direction of arrival (DOA) of desired user, where the phase ambiguity due to  HAD structure  is addressed successfully. Subsequently, a machine-learning (ML) framework is proposed to improve the precision of measuring DOA.  In the training stage, the HAD transceiver works as a receiver and repeatedly measures the values of DOA via I-HAD-ESPRIT to form a slightly large training data set (TDS) of  DOA. From TDS, we find that the probability density function (PDF) of DOA  measurement error (DOAME) is approximated as a Gaussian distribution by the histogram method in ML. This TDS is used to learn the mean of DOA and the variance of DOAME, which are utilized to infer their values and improve their precisions in the real-time stage.  The HAD  transceiver rapidly measures the real-time value of DOA some times to form a relatively small real-time  set  (RTS), which is used to learn the real-time mean and variance of DOA/ DOAME.  Then, three weight combiners are proposed to combine the-maximum-likelihood-learning outputs of TDS and RTS. Their weight factors depend intimately on the numbers of elements in TDS and RTS, and signal-to-noise ratios during the two stages. Using the mean and variance of DOA/DOAME, their PDFs can be given directly, and we propose a robust beamformer for directional modulation (DM) transmitter with HAD by fully exploiting the PDF of DOA/DOAME, especially a robust analog beamformer on RF chain. Simulation results show that: 1) The proposed I-HAD-ESPRIT can achieve the HAD CRLB; 2) The proposed ML framework performs much better than the corresponding real-time one without training stage,  3)  The proposed robust  DM  transmitter  can perform  better  than the corresponding non-robust ones in terms of bit error rate and secrecy rate.
\end{abstract}


%
\IEEEpeerreviewmaketitle

\section{Introduction}
Direction of arrival (DOA) measurement, also known as direction finding, belongs to one of the most important aspects in array signal processing \cite{Godara, Gorodnitsky, Chen, Petre, Zai}. It is also a very old research field and has a long and successful history\cite{Aboulnasr, Xiaofei, Shafin, Liangtian}. In the recent years, massive MIMO promises to provide a ultra-high-resolution DOA estimation \cite{Hongji}. Therefore, the research field of estimating DOA regains its new life.  The need for high resolution DOA estimation always exists  because of its wide applications in many practical scenarios such as radar, sonar, seismic exploration, electronic countermeasures and radio astronomy\cite{Tuncer}. More importantly, with the rapid development of some new technologies and  appearance of new application scenarios such as directional modulation(DM), device to device(D2D), satellite communications, internet of things (IoT), uninhabited aerial vehicle(UAV), 5G and beyond mobile works like  millimeter wave communications, in our view, the demand for DOA estimation will emerge as an explosive growth in the coming future.

Classic and representative DOA estimation methods are  categorized into two main kinds: conventional and subspace methods. Conventional methods, also called classical
methods,  first compute a spatial spectrum and then  find DOAs by numerical search of local maxima of the
spectrum. Typical Methods falling in this class are  Capon and Bartlett methods. But these methods
suffer from  angular resolution loss. Alternatively, two classic high-angular-resolution subspace methods
such as MUSIC (Multiple signal classification) and ESPRIT  (Estimation of signal parameters via rotation invariance techniques) algorithms are more popular and achieve wide applications\cite{Schmidt,Roy}.


 Recently,  compressed sensing(CS) was also applied to improve the precision of DOA estimation. In \cite{Malioutov}, the authors developed an effective sparse representation $\mathrm{L1-SVD}$ algorithm, which first used singular value decomposition(SVD) of data matrix to make a reduction of dimensionality and then the problem of DOA estimation was casted a second-order cone(SOC) optimization. In \cite{Md}, a mixed $l_{2,0}$ approximation DOA algorithm was proposed to solve the joint-sparse recovery problem, which could distinguish closely spaced and highly correlated sources without knowing the number of emitters in advance.

The authors in \cite{Hongji} applied deep learning into massive MIMO system to conduct DOA estimate. Specifically, training data set (TDS) is formed by using randomly generated input and output data, and the weight parameters of deep neural network(DNN) is trained  to learn the statistic property of angle domain. Once DNN is trained well, DNN can output the real-time value of DOA rapidly. In \cite{Qinglong}, an adaptive DOA estimation method, based on convolutional neural network(CNN) with long short term memory, was developed. In \cite{Soumitro}, the authors proposed a CNN-based DOA estimation method for broad band-width signal with the phase of short-time Fourier transform coefficients as inputs of CNN to implement training process.

However, all these DOA related works consider the fully digital (FD) structure, in which each antenna is equipped with one dedicated RF chain. If the number of antennas goes to large-scale, it will lead to a significant increasement in circuit cost and energy consumption. Therefore,  a hybrid analog-to-digital (HAD) MIMO structure is  an excellent alternative solution to FD due to its low circuit cost  and high energy efficiency. The authors in \cite{Qin}  developed a low-complexity HAD Root-MUSIC algorithm in MIMO receive array with phase alignment, which could achieve the CRLB of HAD.

Compared to ESPRIT-based  algorithms, the MUSIC-based methods are more accurate, and stable especially in the high SNR scenarios\cite{Robert}. But, the need for gratuitous subarray calibration information brings more inconveniences for MUSIC-based ones owing to that the phase drifts will reduce the reliability of received data\cite{Sidiropoulos}. More important, for Root-MUSIC-based methods, to resolve multiple emitters,  the receiver should know the number of incident waves in advance while ESPRIT doesn't need such a number and  is even  capable of  estimating this number \cite{Roy}.

In this paper, to avoid the disadvantages of MUSIC-based algorithms, an  ESPRIT-based  DOA estimator is proposed for a HAD receiver. Here, the phase ambiguity is addressed by maximizing the output receive power, which is similar to the Root-MUSIC in \cite{Qin}. However, the kind of methods will face a performance loss. With the help of machine learning framework,  we further improve the performance of this kind of method.  Subsequently, to obtain more precise variance, we make a further investigation of the statistical feature of DOAMEs and conduct a weight combination of the DOA means and variances of TDS and new on-line data set or real-time set (RTS). Since we complete the density estimation of DOA, we propose a robust beamforming scheme for DM using HAD structure. Before we summarize  our  main contributions, let us review the literature concerning DM as follows.

 As a secure physical-layer transmission technology, DM  need to know the angle direction information of desired user in advance, which can be obtained by the preceding methods.  The existing research works about DM synthesis can be divided into two categories. The first one was  based on RF front-end\cite{Baba,Daly,Shi}. In \cite{Baba}, the authors developed a near-field direct antenna modulation technique using a large amount of reflectors and switches to adjust the amplitude and phase of signal, thus distorting  signal constellation severely. In \cite{Daly}, the authors proposed a phased-array-based DM,  which produced phase and amplitude of each symbol in desired direction by shifting phase of each array element. Similarly, in \cite{Shi},  an improved enhancing direct antenna modulation was presented by increasing element spacing to improve the direction error rate in desired direction. The second way of DM synthesis is implemented in baseband with the aid of artificial noise (AN) to disturb the constellation pattern along undesired directions. In \cite{Ding},  the orthogonal vector method with the help of AN was proposed to synthesize the waveform of DM. In \cite{Hu},  a closed-form expression of null-space-projection(NSP) beamformer was derived  and a robust DM synthesis was proposed using the rule of conditional minimum mean square error for single-desired-user scenario. Furthermore, in \cite{Wu,Zhu}, the concept of robust DM is extended to  two new scenarios: multi-beaming broadcasting and MU-MIMO, respectively. In \cite{Zhu2,Zhou}, the relay-aided DM is exploited to enhance the physical layer security. In \cite{Zhu2,Hu2,Wu2}, the authors proposed a new secure concept, called secure precise wireless transmission (SPWT). The former achieved the SPWT with the help of multi-relay  while the latter  adopted a random frequency diverse array plus DM and random subcarrier selection plus DM with the help of AN, respectively. In \cite{Wan}, the optimal power allocation strategy between confidential message and AN was optimized to maximize the secrecy rate (SR) of DM system. An alternative iteration method to maximize SR by resorting to GPI was proposed in \cite{Yuhai}.  Additionally, DM is utilized to harvest energy by its directional property \cite{Zhou} due to its excellent directive property. In particular, in a massive MIMO scenario, an extremely high energy-efficiency and security for harvesting energy and transmitting confidential messages can be achieved. From this aspect, we will predict the fact that DM will become a dramatically important secure wireless transmission way in the future wireless networks.

However, all the above research works concerning DM focus on fully-digital transmitter, if a massive transmit antenna array is adopted, a hybrid analog and digital DM transmitter is a perfect choice of striking a good balance by  taking circuit cost, energy consumption, and computation complexity into account. Traditional research works pertaining HAD architectures focus on the design of two beamforming matrices including digital precoder and analog precoder at transmitter. In \cite{Xinying}, the authors  first introduced HAD  and verified the number of RF chains needed to realize the FD precoding in single-data-stream scenario. In \cite{Foad1, Foad2}, the HAD is generalized to multi-user MIMO scenarios. In \cite{Xianghao},  an effective alternating minimization algorithms was proposed for both fully-connected and sub-connected  HAD architectures. In \cite{Xinyu},  a energy-efficient precoder based on successive interference cancelation was conceived for  partially-connected architecture. The authors in \cite{Yahia} considered security in HAD systems by maximizing the projection between FD precoder and the hybrid precoder. In \cite{Heath2016An},  an overview of hybrid precoding techniques especially in mmWave MIMO systems was presented. However, all the preceding works optimized the two beamformers: analog and digital,  without   AN projection.

To the best of our knowledge, there is still no  scheme of combining HAD and DM to provide a robust secure transmission with low-complexity and low cost in line-of-sight (LoS) channels. Therefore, in this paper, a robust HAD plus DM transmitter is presented with the aid of AN to achieve a robust and secure physical-layer transmission by exploiting the PDF of measured DOA. Our main contributions are summarized as follows:

\begin{enumerate}
 \item  A  ESPRIT DOA measurement method for a sub-connected receiver with HAD structure  is proposed  to estimate directions of desired and undesired users, respectively. By utilizing the rotational invariance among each subarray in sub-connected hybrid architecture, the ESPRIT is well applicable to hybrid analog and digital sub-connected receiver. And the phase ambiguity due to antenna subarray structure  is addressed by maximizing the output power over a set of multiple potential solutions induced by  periodicity of phase. Subsequently, after applying the proposed ESPRIT repeatedly to form a set of measured values of DOA,   via histogram method in machine learning (ML), we find that the density distribution of DOA or its measurement error (DOAME) can be approximated as a Gaussian distribution, which significantly simplifies the following ML-based improved method of DOA estimation.
 \item To improve the estimate performance of ESPRIT, a ML-based framework is proposed. In this framework, in the first stage, called training or offline stage,  ESPRIT is repeatedly used to form a TDS, which will provide a high-precision estimate for the mean of DOA and variance of DOAME. Here, the number of  elements in TDS can be chosen to sufficiently large in order to achieve a high-performance predictor for variance of DOAME and mean of DOA in the next stage.  In the second stage, called real-time or new-data, the value of DOA are rapidly estimated by several snapshots,  and yields the mean of real-time DOA and a coarse estimate of DOAME variances by using maximum likelihood learning. Finally, the outcomes from the above stages are combined to form new outputs by a weight method, where the weighted factors rely heavily on the receive SNR, and the numbers of TDS and RTS.  Now, we can directly have the probability density function (PDF) of DOAME. The above process actually is the density estimation.  From simulation results, it follows that, compared to the corresponding method without training stage, the proposed ML method can dramatically enhance the performance of DOA  and DOAME variance in terms of variance. This paves a way to design a robust beamforming method of DM for HAD-MIMO structure.

 \item After learning the mean values of DOAs and PDFs of DOAMEs of desired user and eavesdropper,  we can start to design a robust beamformer for DM in HAD MIMO system, which is made up of three beamforming vectors/matrices required to be devised: digital precoding vector of confidential message in baseband, digital AN projection matrix in baseband, and analog beamforming matrix on RF chain. The first two beamformers are optimized  via minimizing the Euclidean distance between the HAD beamforming matrix and  corresponding FD precoder, respectively. Our emphasis is mainly on  the last  analog beamforming matrix. By making use of conditional expectation, a robust analog beamforming matrix is proposed. Simulation results reveal that, compared to non-robust beamforming methods,  our proposed robust hybrid method performs much better in terms of both SR and BER.
\end{enumerate}

The remainder of this paper is organized as follows. Section II describes the system models of HAD receiver to learn DOA and HAD DM transmitter to achieve a robust secure transmission. In Section III,  the  ESPRIT DOA estimation using HAD structure is proposed,  the density of DOAMEs is approximately attained  by histogram method,  and the TDSs of DOA measurements are constructed.  Finally,  a weight combination between the outputs of TDS and RTS is to yield a more  precise mean and  variance of DOA.  In Section IV, based on conditional expectation,  a robust secure beamformer is proposed for DM HAD transmitter with emphasis on analog part. Simulation results will be presented in Section V. Finally, we draw our conclusions  in Section VI.

Notations: throughout the paper, matrices, vectors, and scalars are denoted by letters of bold upper case, bold lower case, and lower case, respectively. $(\cdot)^T$, $(\cdot)^*$£¬ and $(\cdot)^H$ denote transpose,~conjugate,~and conjugate transpose,~respectively. $\parallel\cdot\parallel_2$ and $\parallel\cdot\parallel_F$ denote the $l_2$ norm of a vector and Frobenius norm of a matrix, respectively. $\text{tr}(\cdot)$ and $vec(\cdot)$ are matrix trace and matrix vectorization. $\otimes$ denotes the Kronecker products between two matrices.

\section{System Model}
\begin{figure}[htp]
\centering
\subfigure[Hybrid DOA estimation]{
\includegraphics[width=8.4cm]{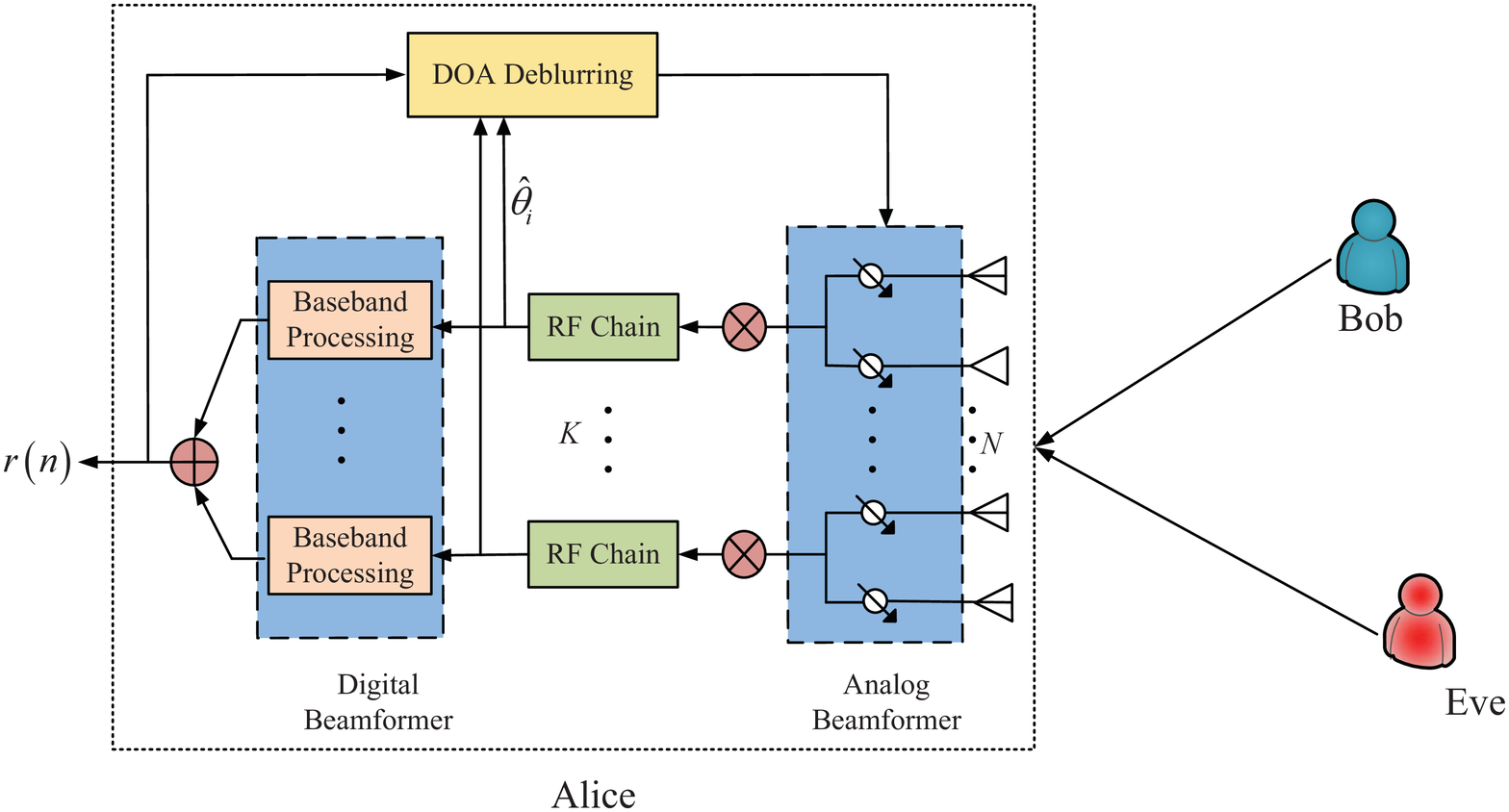}}
\subfigure[Hybrid DM precoder]{
\includegraphics[width=8cm]{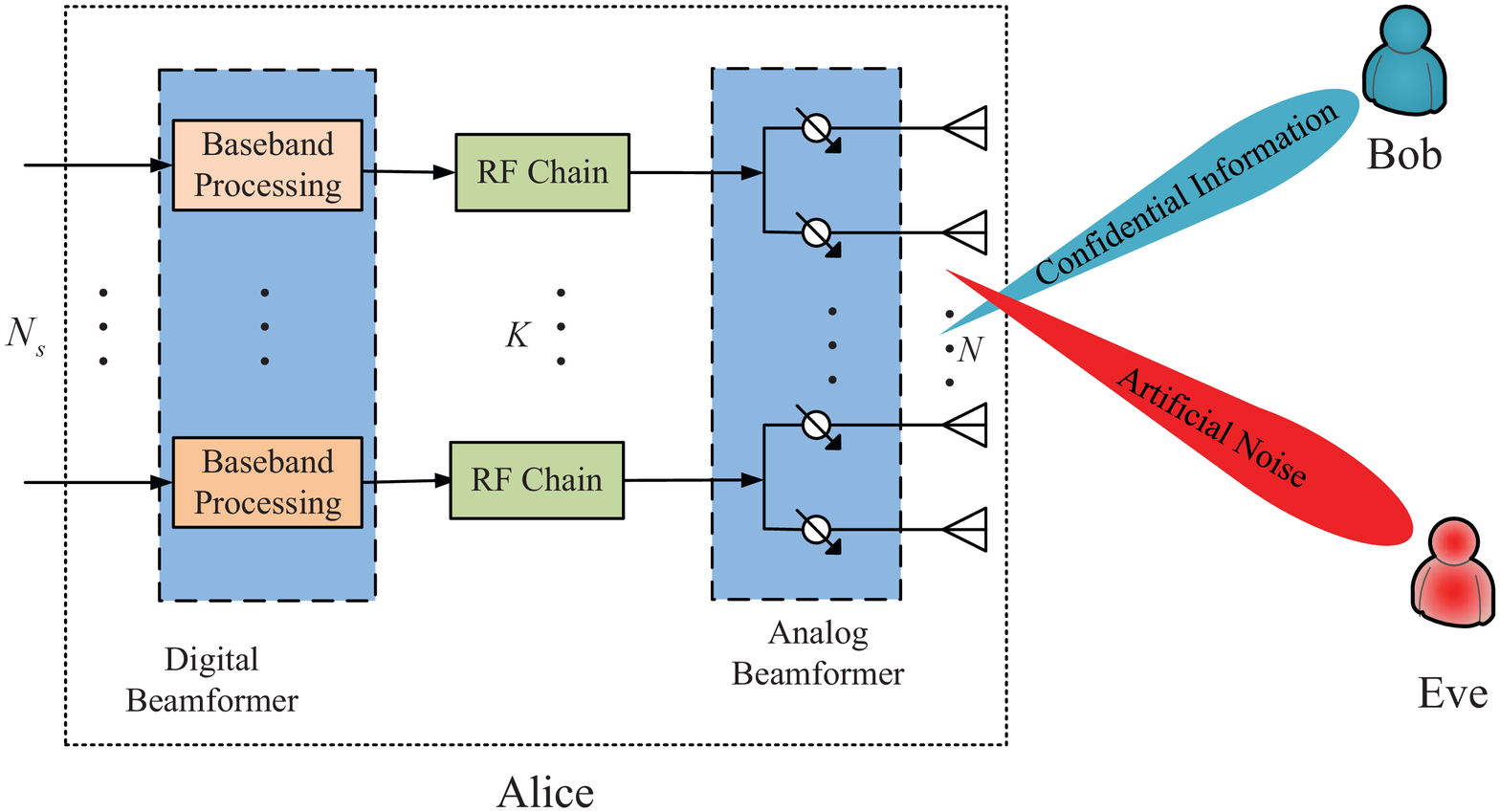}}
\caption{Block diagram of DOA measurement and DM for HAD transceiver}
\label{system}
\end{figure}
In Fig.~\ref{system}, the schematic diagram of a transceiver using HAD is plotted. In the figure, there are two subfigures (a) and (b). They show the fact that the  working process of the practical DM  system consists of two steps. Before performing the beamforming operation of DM, we should measure the directional angle of desired user or its statistical property as shown in subfigure (a). Subsequently, in subfigure  (b), using the measured value of DOA or its statistical property, the beamforming scheme can be designed. In both subfigures, the sub-connected hybrid architecture is adopted, where each array is made up of $K$ sub-arrays of antennas and each sub-array  has $M$ antennas. In other words,  $N=KM$, where $N$ is the total number of antennas at Alice.  Below, we establish the system models of subfigures (a) and (b).

Fig.~\ref{system} (a) is a HAD receiver at Alice. In this subfigure, a narrow-band signal $s(t)e^{jw_ct}$  from a far-field emitter Bob impinges on the HAD receive array at Alice,  the receive signal is
\begin{align}\label{Rx-S-AB}
\mathbf{y}_a(t)=\mathbf{F}_{RF}^H\mathbf{a}(\theta)s(t)e^{jw_ct}+\mathbf{n}(t)
\end{align}
where  $s(t)$ is the baseband signal, $w_c=2\pi f_c$ with $f_c$ being carrier frequency, and $\mathbf{a}(\theta)$ is the array manifold defined as
\begin{align}
\mathbf{a}(\theta)=[1,e^{j\frac{2\pi}{\lambda}  d\cos\theta},\cdots,e^{j\frac{2\pi}{\lambda}(N-1)d\cos\theta}]^T
\end{align}
which is used to estimate the angle range from $0^\circ$ to $180^\circ$  and convenient to design beamformer for DM in the following sections. In (\ref{Rx-S-AB}), $\mathbf{F}_{RF}$ is the $N\times K$ analog receive beamforming matrix consisting of $K$ analog weight vectors $\mathbf{f}_{k}$, with each element having the same amplitude $\frac{1}{\sqrt{M}}$ but different phases, as follows
\begin{align}
\mathbf{F}_{RF}=\left[ {\begin{array}{*{20}{c}}
{\mathbf{f}_1}&{\mathbf{0}}&{\cdots}&{\mathbf{0}}\\
{\mathbf{0}}&{\mathbf{f}_2}&{\cdots}&{\mathbf{0}}\\
{\vdots}&{\vdots}&{\ddots}&{\vdots}\\
{\mathbf{0}}&{\mathbf{0}}&{\cdots}&{\mathbf{f}_K}
\end{array}} \right]
\end{align}
where $\mathbf{f}_k=\frac{1}{\sqrt{M}}\left[e^{j\alpha_{k,1}},\cdots,e^{j\alpha_{k,M}}\right]^T$.
Then, after performing the down-conversion and analog-to-digital conversion (ADC) operations, the receive signal at Alice can be expressed as
\begin{align}
\mathbf{y}_a(n,\theta)=\mathbf{F}_{RF}^H\mathbf{a}(\theta)s(n)+\mathbf{n}
\end{align}
Finally, after the digital beamforming operation, the receive signal can be written as
\begin{align}
r(n)=\mathbf{F}_{BB}^H\mathbf{F}_{RF}^H\mathbf{a}(\theta)s(n)+\mathbf{F}_{BB}^H\mathbf{n}
\end{align}
By making use of the above model, we will develop an improved ESPRIT DOA estimator for HAD architecture to learn the direction angle of desired user. The specific DOA estimation process will be presented in Section III.

After obtaining the value of DOA of Bob, now we show how to implement a DM transmitter using HAD beamforming as  indicated in Fig.~\ref{system} (b). Here, Alice works on transmit model. The DM transmitter  is equipped  with $N$ uniformly spaced linear antennas array. The transmit signal is
\begin{align}
\mathbf{s}=\sqrt{\beta P_s}\mathbf{V}_{RF}\mathbf{v}_{BB}x+\sqrt{(1-\beta)P_s}\mathbf{V}_{RF}\mathbf{T}_{BB}\mathbf{z}
\end{align}
where $P_s$ is the total transmit power constraint, $\beta$ denotes the power allocation (PA) factor of confidential messages, and $1-\beta$ stands for the PA factor of AN. Additionally, $\mathbf{V}_{RF}\in \mathbb{C}^{N\times K}$ and $\mathbf{v}_D\in \mathbb{C}^{K\times 1}$ represent the analog and digital beamforming martices of carrying confidential messages to desired user, respectively. $\mathbf{V}_{RF}$ has the  following form
\begin{align}\label{V-RF}
\mathbf{V}_{RF}=\left[ {\begin{array}{*{20}{c}}
{\mathbf{v}_1}&{\mathbf{0}}&{\cdots}&{\mathbf{0}}\\
{\mathbf{0}}&{\mathbf{v}_2}&{\cdots}&{\mathbf{0}}\\
{\vdots}&{\vdots}&{\ddots}&{\vdots}\\
{\mathbf{0}}&{\mathbf{0}}&{\cdots}&{\mathbf{v}_K}
\end{array}} \right]
\end{align}
where $\mathbf{v}_k$ is the analog beamforming vector of subarray $k$,  $\mathbf{T}_{BB}\in \mathbb{C}^{K\times N_s}$ denotes the digital AN projection matrices with constraints $\parallel\mathbf{V}_{RF}\mathbf{T}_{BB}\parallel^2_F=\parallel\mathbf{V}_{RF}\mathbf{v}_{BB}\parallel^2=1$. $x\sim \mathcal{CN}(0,1)$ and $\mathbf{z}\sim \mathcal{CN}(0,\mathbf{I})$ are the confidential message and AN interfering the eavesdropper, respectively. Let us define $\Omega({\theta})=\frac{2\pi d}{\lambda}\cos({\theta})$ and the total steering vector (SV) in free space as
\begin{align}
\mathbf{h}^H(\theta)=[&e^{j(1-\frac{N+1}{2})\frac{2\pi d}{\lambda}\cos(\theta)},e^{j(2-\frac{N+1}{2})\frac{2\pi d}{\lambda}\cos(\theta)},\cdots,\nonumber\\
&e^{j(N-\frac{N+1}{2})\frac{2\pi d}{\lambda}\cos(\theta)}]=[\mathbf{h}_1^H(\theta),\cdots,\mathbf{h}_K^H(\theta)].
\end{align}
where $\mathbf{h}_k(\theta)^H$ corresponds to the steering vector of subarray $k$ given by
\begin{align}\label{Subarray-k-SV}
\mathbf{h}_k(\theta)^H=\frac{1}{\sqrt{M}}\left[e^{j\alpha_{k,1}},\cdots,
e^{j\alpha_{k,M}}\right]
\end{align}
where
\begin{align}
\alpha_{k,m}=\left[(k-1)M+m-\frac{N+1}{2}\right]\Omega(\theta).
\end{align}
Interestingly, the total steering vector $\mathbf{h}^H(\theta)$ is also written in the following form
\begin{align}
\mathbf{h}^H(\theta)=\mathbf{h}^H_\Theta(\theta)\otimes \mathbf{h}^H_1(\theta)
\end{align}
where $\mathbf{h}^H_\Theta(\theta)$ is defined as
\begin{align}
\mathbf{h}^H_\Theta(\theta)=\left[e^{j0},~e^{jM\Omega(\theta)},~\cdots,
e^{-j(K-1)M\Omega(\theta)}\right].
\end{align}
The receive signal at the desired user Bob can be expressed as
\begin{align}
y(\theta_d)&=\mathbf{h}^H(\theta_d)\mathbf{s}+n_d \nonumber\\
&=\sqrt{\beta P_s}\mathbf{h}^H(\theta_d)\mathbf{V}_{RF}\mathbf{v}_{BB}x\nonumber\\
&+\sqrt{(1-\beta)P_s}\mathbf{h}^H(\theta_d)\mathbf{V}_{RF}\mathbf{T}_{BB}\mathbf{z}+n_d
\end{align}
where
$n_d$ is the complex additive white Gaussian noise(AWGN) with $n_d\sim\mathcal{C}\mathcal{N}(0, \sigma_d^2)$. Similarly, the receive signal at eavesdropper is
\begin{align}
y(\theta_e)&=\mathbf{h}^H(\theta_e)\mathbf{s}+n_e \nonumber\\
&=\sqrt{\beta P_s}\mathbf{h}^H(\theta_e)\mathbf{V}_{RF}\mathbf{v}_{BB}x \nonumber\\
&+\sqrt{(1-\beta)P_s}\mathbf{h}^H(\theta_e)\mathbf{V}_{RF}\mathbf{T}_{BB}\mathbf{z}+n_e
\end{align}
where
$n_e$ is the complex AWGN following $n_e\sim\mathcal{C}\mathcal{N}(0, \sigma_e^2)$. In particular, it is noted that Eve is assumed to be a passive eavesdropper in our paper. According to (7), the achievable rate of the desired user Bob is
\begin{align}
R(\theta_d)=log_2(1+\frac{\beta P_s\parallel\mathbf{h}^H(\theta_d)\mathbf{V}_{RF}\mathbf{v}_{BB}\parallel^2}{(1-\beta)P_s\parallel\mathbf{h}^H(\theta_d)\mathbf{V}_{RF}\mathbf{T}_{BB}\parallel^2+\sigma^2})
\end{align}
Similarly, the achievable rate of eavesdropper is
\begin{align}
R(\theta_e)=log_2(1+\frac{\beta P_s\parallel\mathbf{h}^H(\theta_e)\mathbf{V}_{RF}\mathbf{v}_{BB}\parallel^2}{(1-\beta)P_s\parallel\mathbf{h}^H(\theta_e)\mathbf{V}_{RF}\mathbf{T}_{BB}\parallel^2+\sigma^2})
\end{align}
Then the SR can be defined as the rate difference between $R(\theta_b)$ and $R(\theta_e)$
\begin{align}
R_S=\max\{0,R(\theta_d)-R(\theta_e)\}.
\end{align}

\section{Proposed ML-based framework for  DOA estimation using HAD structure}
In this section, when classic ESPRIT is applied to HAD receiver, there exists a serious problem of phase ambiguity. By maximizing the receive power over  a set of finite potential angles, we completely solve this problem to form an improved  ESPRIT. Using this improved  ESPRIT and histogram  method in ML,  we find an important fact: the DOAME obeys a Gaussian distribution.  Then, TDS is constructed.   Via ML framework, the value of DOA and its density is inferred from this TDS and the following RTS.

\subsection{Improved ESPRIT-based HAD DOA Estimator}
First, let us assume the phases of all elements of analog beamforming matrix are equal to zeros, that is, each analog subarray satisfies
\begin{align}
\mathbf{f}_k=\frac{1}{\sqrt{M}}\left[1,\cdots,1\right]^T.
\end{align}
Then, in  accordance with  (4), we can directly obtain the output vector after analog beamforming as follows
\begin{align}
&\mathbf{y}_a(n,\theta)=\mathbf{F}_{RF}^H\mathbf{a}(\theta)s(n)+\mathbf{n}\nonumber\\
&=\frac{1}{\sqrt{M}}\left[{\begin{array}{*{20}{c}}
{1\cdots1}&{\mathbf{0}}&{\cdots}&{\mathbf{0}}\\
{\mathbf{0}}&{1\cdots1}&{\cdots}&{\mathbf{0}}\\
{\vdots}&{\vdots}&{\ddots}&{\vdots}\\
{\mathbf{0}}&{\mathbf{0}}&{\cdots}&{1\cdots1}
\end{array}}\right]\mathbf{a}(\theta)s(n)+\mathbf{n}\nonumber\nonumber\\
&=\frac{1}{\sqrt{M}}f(\theta)\mathbf{a}_K(\theta)s(n)+\mathbf{n}
\end{align}
where
\begin{align}
f(\theta)=\sum_{m=0}^{M-1}e^{j\frac{2\pi}{\lambda}md\cos\theta}
=\frac{1-e^{j2\pi/\lambda Md\cos\theta}}{1-e^{j2\pi/\lambda d\cos\theta}},
\end{align}
and
\begin{align}
\mathbf{a}_K(\theta)=\left[1,e^{j\frac{2\pi}{\lambda}Md\cos\theta},\cdots,e^{j\frac{2\pi}{\lambda}(K-1)Md\cos\theta}\right]^T.
\end{align}
Let $\mathbf{a}_D(\theta)=\frac{1}{\sqrt{M}}f(\theta)\mathbf{a}_K(\theta)$, then $\mathbf{y}_a(n)$ can be recasted as
\begin{align}
\mathbf{y}_a(n)=\mathbf{a}_D(\theta)s(n)+\mathbf{n}.
\end{align}
Next, we adopt ESPRIT to estimate the directional angle of emitter. We choose the first $K-1$ sub-array of antennas to be Array 1, and choose the last $K-1$ sub-array of antennas to be Array 2, then the output can be expressed as
\begin{align}
\mathbf{y}_{a}^1(n,~\theta)=\mathbf{a}_{D}^1(\theta)s(n)+\mathbf{n}^1,
\end{align}
and
\begin{align}
\mathbf{y}_{a}^2(n)&=\mathbf{a}_{D}^2(\theta)s(n)+\mathbf{n}^2\nonumber\\
&=\mathbf{a}_{D}^2(\theta)\mathbf{\Phi}s(n)+\mathbf{n}^2,
\end{align}
respectively,  where
\begin{align}
\mathbf{a}_{D1}(\theta)=\frac{f(\theta)}{\sqrt{M}}[1,e^{j\frac{2\pi}{\lambda}Md\cos\theta},\cdots,e^{j\frac{2\pi}{\lambda}(K-2)Md\cos\theta}]^T
\end{align}
and $\mathbf{\Phi}=\text{diag}\{e^{j\frac{2\pi}{\lambda}Md\cos\theta}\}$. By computing the co-variance matrix of $\mathbf{y}_{a}^1(n)$ and $\mathbf{y}_{a}^2(n)$, separately, we can obtain the signal spaces $\mathbf{E}_1$ and $\mathbf{E}_2$. Because of the translational displacement, there must exit a non-singular transformation matrix $\mathbf{\Psi}$  such that
\begin{align}
\mathbf{E}_1\mathbf{\Psi}=\mathbf{E}_2,
\end{align}
and there must exit a non-singular transformation matrix $\mathbf{T}$ such that
\begin{align}
\mathbf{E}_1=\mathbf{a}_{D1}\mathbf{T},
\end{align}
and
\begin{align}
\mathbf{E}_2=\mathbf{a}_{D1}\mathbf{\Phi}\mathbf{T}.
\end{align}
According to Eqs. (13),~(14)~and (15), we can derive a crucial relationship
\begin{align}
\mathbf{T}\mathbf{\Psi}\mathbf{T}^{-1}=\mathbf{\Phi}.
\end{align}
Therefore, it is evident that the eigenvalues of $\mathbf{\Psi}$ are equal to the diagonal elements of $\mathbf{\Phi}$, i.e.
\begin{align}
\lambda_i=\exp\left\{j\frac{2\pi}{\lambda}Md\cos\theta\right\}.
\label{26}
\end{align}
Now the problem is how to calculate the rotation operator $\mathbf{\Psi}$. Next, we use the total least-squares in \cite{Roy} method to estimate $\mathbf{\Psi}$.
The correlation matrix of the entire antenna array is
\begin{align}\label{Ryy}
\mathbf{R}_{yy}&=E[\mathbf{y}_{a}(n)\mathbf{y}_a^H(n)]\approx\frac{1}{L}\{\mathbf{Y}_a\mathbf{Y}_a^H\}\nonumber\\
&\approx\mathbf{a}_D\sigma_s^2P_b^2\mathbf{a}_D^H+\sigma_n^2\mathbf{I}_K
\end{align}
where $\mathbf{Y}_{a}=[\mathbf{y}_{a}(1),\mathbf{y}_{a}(2),\cdots,\mathbf{y}_{a}(L)]$ is the matrix of spatial-time sampling points. The correlation matrix of the entire antenna array has $K$ eigenvalues, and the associated $K$  eigenvectors $\mathbf{E}=[\mathbf{e}_1 \mathbf{e}_2 \cdots \mathbf{e}_K]$. Signal subspace $\mathbf{E}_s$ consists of the eigenvectors corresponding to the largest eigenvalues of the matrix $\mathbf{R}_{yy}$.  Since signal spaces $\mathbf{E}_1$ and $\mathbf{E}_2$ can be constructed from the entire antennas array signal space $\mathbf{E}_s$, we choose the first $K-1$ rows of $\mathbf{E}_s$ to construct $\mathbf{E}_1$, and similarly, the last $K-1$ rows to compose $\mathbf{E}_2$. Next,
we utilize signal subspaces to construct a $2\times2$ matrix.
\begin{align}
\mathbf{C}=\left[ {\begin{array}{*{20}{c}}
{\mathbf{E}_1^H}\\{\mathbf{E}_2^H}\\\end{array}} \right]\left[\mathbf{E}_1 \mathbf{E}_2\right]
=\mathbf{E}_C\Lambda\mathbf{E}_C^H
\end{align}
where $\mathbf{E}_c$ can be derived from the EVD (Eigen value decomposition) of matrix $\mathbf{C}$ and $\Lambda=\text{diag}\{\lambda_1,\lambda_2\}$.
Furthermore, $\mathbf{E}_c$ can be decomposed into four subarrays as follows
\begin{align}
\mathbf{E}_c=\left[ {\begin{array}{*{20}{c}}
{\mathbf{E}_{11}}&{\mathbf{E}_{12}}\\{\mathbf{E}_{21}}&{\mathbf{E}_{22}}\\\end{array}} \right]
\end{align}
Then, we can obtain the rotation operator $\mathbf{\Psi}$
\begin{align}
\mathbf{\Psi}=-\mathbf{E}_{12}\mathbf{E}_{22}^{-1}
\end{align}
By calculating the eigenvalue $\lambda_1$ of $\mathbf{\Psi}$, the angle of emitter estimated by  traditional ESPRIT method is
\begin{align}
\hat{\theta}=\arccos(\frac{\lambda\arg(\lambda_1)}{2\pi Md})
\label{31}
\end{align}
However, the periodicity in equation (\ref{26}) generates an effect of phase blurring, which makes estimated angle no longer the expression in equation(\ref{31}), thus, we need to extend the estimated angle to a set of potential solutions
\begin{align}
\hat{\Theta}=\{\hat{\theta_i},i\in{0,1,\cdots,M-1}\}
\end{align}
where
\begin{align}
\hat{\theta_i}=\arccos(\frac{\lambda(arg(\lambda_1)+2\pi i)}{2\pi Md})
\end{align}
Then, computing the receive signals power corresponding to the $M$ angles individually, we choose the angle of maximizing the output power as the optimal value casted as
\begin{equation}
\hat{\theta}=\mathop{\arg\min}_{\hat{\theta_i}\in\hat{\Theta} } \ \  P_r(\hat{\theta_i})=\frac{1}{LN^2}\sum_{n=1}^L(r(n)r(n)^H)
\end{equation}
where
\begin{align}
r(n)&=\mathbf{F}_{BB}^Hy_a(n)\nonumber\\
&=\sum_{k=1}^Ke^{-j\alpha_k}y_a(n)\nonumber\\
&=\frac{1}{\sqrt{M}}s(n)\sum_{k=1}^Ke^{j\frac{2\pi}{\lambda}(k-1)Md(\cos\theta_0-\cos\hat{\theta}_i)}\nonumber\\
&\times\sum_{m=1}^Me^{j\frac{2\pi}{\lambda}(m-1)d(\cos\theta_0-\cos\hat{\theta}_i)}+\sum_{k=1}^Ke^{-j\frac{2\pi}{\lambda}(k-1)Md\cos\hat{\theta_i}}n
\end{align}

 Since we have obtained  the value of DOA  of incoming signal, we can  further estimate the SNR of incoming signal. Given the estimated angle $\hat{\theta}$, the analog beamforming vector (ABV) of each subarray can be designed as
\begin{align}
\hat{\mathbf{f}}_k=\frac{1}{\sqrt{M}}[e^{j\frac{2\pi}{\lambda}((k-1)M+1)d\cos\hat{\theta}},\cdots,e^{j\frac{2\pi}{\lambda}((k-1)M+M)d\cos\hat{\theta}}]^T
\end{align}
Then, after the ABV $\hat{\mathbf{f}}_k$  is performed in RF chain, the output signal of subarray $k$ in the next time slot  can be given by
\begin{align}\label{output-subk}
\hat{y}_{a,k}(n)=\hat{\mathbf{f}}_k^H\mathbf{a}_k(\theta_0)s(n)+n.
\end{align}
If the estimated angle $\hat{\theta}$ is equal to the exact angle $\theta_0$, then $\hat{\mathbf{f}}_k^H\mathbf{a}_k(\theta_0)=1$, and (\ref{output-subk}) can be recasted as
\begin{align}
\hat{y}_{a,k}(n)=s(n)+n
\end{align}
which yields the output vector of  all subarrays
\begin{align}
\hat{\mathbf{y}}_a(n)=\left[\hat{y}_{a,1}(n),\hat{y}_{a,2}(n),\cdots,\hat{y}_{a,K}(n)\right]^T.
\end{align}
Then we have the overall estimated SNR $\hat{\rho}$ in HAD architecture as follows
\begin{align}
\hat{\rho}=\frac{\hat{\mathbf{y}}_a(n)^T\hat{\mathbf{y}}_a(n)^*-K\sigma_n^2}{K\sigma_n^2}.
\end{align}
where $\sigma_n^2$ is taken to be  the least eigen-value of matrix (\ref{Ryy}). Until now, we complete the DOA and SNR estimate by the improved ESPRIT, called improved HAD-ESPRIT (I-HAD-ESPRIT) in what follows.

\subsection{Proposed ML-based  DOA and its density estimation}
 Due to the effect of channel noise or co-channel signals, there exists  DOA measurement error in the estimated DOA in the preceding subsection.  Considering DOAME, the estimated angle can be  modeled as
\begin{align}
\tilde\theta(n)=\theta_0+\Delta\theta(n),~~n=1,\cdots,N_{DOA}
\end{align}
where $\tilde\theta(n)$ represents the practical observation angle, $\theta$ is the ideal angle we want to obtain, and $\Delta\theta(n)$ is the measurement error between the ideal angle and measured one, $n=1,2,\cdots,~N$ denotes the number of time slots. Here, the proposed I-HAD-ESPRIT is used one time to measure DOA per time slot.
\begin{figure}[h]
  \centering
  \includegraphics[width=9cm]{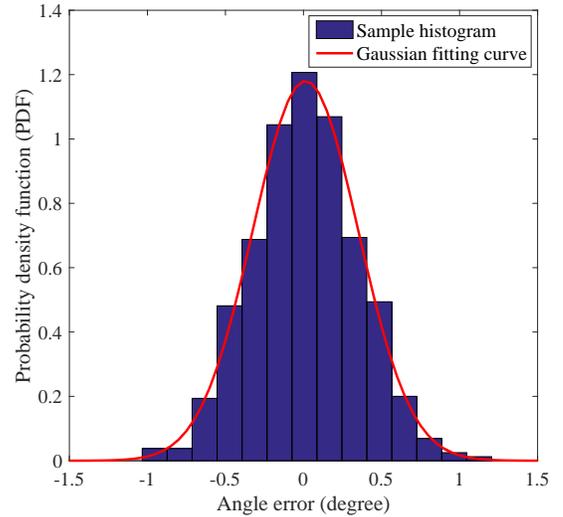}\\
  \caption{DOA measurement error}
  \label{2}
\end{figure}

\begin{figure}[h]
  \centering
  \includegraphics[width=9cm]{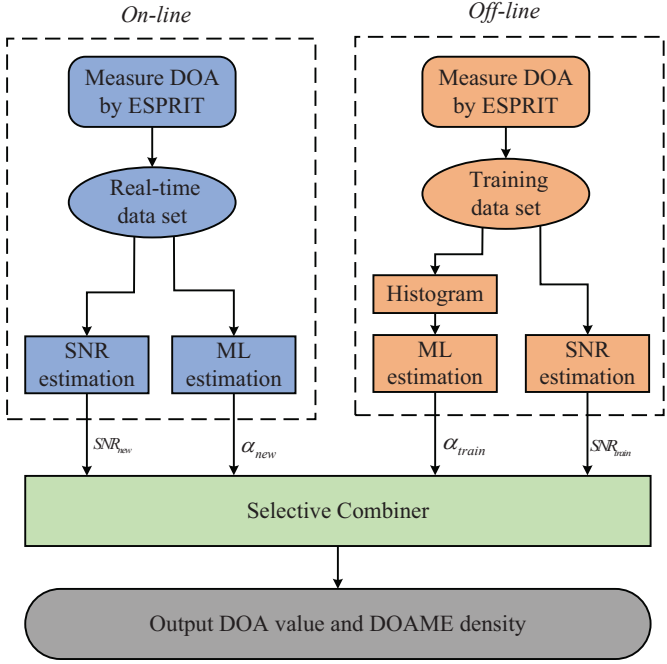}\\
  \caption{Flow diagram of ML-based DOA estimator}
  \label{BL}
\end{figure}

Using the proposed I-HAD-ESPRIT in the preceding subsection repeatedly for several hundreds or thousands, Fig.~\ref{2} depicts the histogram of DOAME $\Delta\theta(n)$ provided that $N=32$,~$M=4$, and SNR=0dB. From this figure, we can find the important fact that the measured value of DOA approximately obeys a Gaussian distribution with zero mean.
 \begin{align}\label{DOA-PDF}
f\left(\tilde\theta(n);\theta,\sigma^2\right)=\frac{1}{\sqrt{2\pi\sigma^2}}\exp\{-\frac{1}{2\sigma^2}(\tilde\theta(n)-\theta)^2\}
\end{align}
If we can learn its variance, then we can give its density function. This is just the density estimation in ML field. Below, we will show how to infer the density of DOAME and improve the accuracy of DOA measurement  by making use of ML-based framework.

In practice, a rapid real-time or on-line DOA measurement is required in order to quickly locate the position of emitter, especially in military field. Thus, it is not allowable to take too many time slots.  This means that the offline  training is very important to infer the density of DOAME and improve the accuracy of DOA measurement. First, the TDS of DOA measurements is offline constructed by using the proposed I-HAD-ESPRIT in the previous subsection. Let us define the TDS as $S_{TDS}=\left\{\tilde{\theta}_1, \tilde{\theta}_{2},\cdots,\tilde{\theta}_{N_{ TDS}}\right\}$, where $N_{NDS}$ stands for the number of elements in TDS.

%
Considering the DOA measurement is approximated as a Gaussian distribution and in terms of the learning rule of maximum likelihood in \cite{Bishop}, we have its mean
\begin{align}
\hat{\theta}_{TDS}=\frac{1}{N_{TDS}}\sum_{\tilde{\theta}_i\in S_{TDS}}\tilde{\theta}_i,
\end{align}
and sampling variance
\begin{align}
\hat{\sigma}^2_{TDS}=\frac{1}{N_{TDS}-1}\sum_{i=1}^{N_{TDS}}(\tilde\theta_i-\hat\theta_{TDS})^2
\end{align}
in the training stage. Similarly, in the real-time measuring stage, we have
\begin{align}
\hat{\theta}_{RTS}=\frac{1}{N_{RTS}}\sum_{\tilde{\theta}_i\in S_{RTS}}\tilde{\theta}_i,
\end{align}
and real-time variance
\begin{align}
\hat\sigma^2_{RTS}=\frac{1}{N_{RTS}-1}\sum_{i=1}^{N_{RTS}}(\tilde\theta_i-{\hat{\theta}}_{RTS})^2.
\end{align}
If Bob keeps fixed or moves away from Alice along the fixed desired direction during both offline and online periods, then we call this scenario a stable state. Clearly,   the above two estimated values mean and variance can be exploited to  improve the accuracy of DOA measurement mean and variance. Conversely, if Bob  moves fast away from the desired direction, we call this situation as a moving state. In such a situation,  at least the estimated variance during the training stage can be used to improve the accuracy of DOA measurement variance by predicting the variance of real-time measuring stage. In what follows, we discuss the two situations. In the stable state, Alice doesn't need to adjust its beamforming direction. However, in the moving state, Alice is required to adjust its beam direction to follow the real-time varying angle of the desired user.

 Due to the two important facts: the SNR of training stage may be different from the SNR of real-time measuring stage, and the number $N_{TDS}$ of elements in TDS is also possibly different from that $N_{RTS}$  in RTS, we should take the two factors into account when we combine the learning output of the two sets. In general, $N_{TDS}>> N_{RTS}$.   In the stable state, the weight output of means and variances due to TDS and RTS are given by
 \begin{align}\label{ML-DOA-ConWei}
\hat\theta_{ML}=\alpha_{1}{\hat\theta}_{TDS}+\alpha_{2}\hat\theta_{RTS}=\theta+\delta\theta_{ML}
\end{align}
with the constraint $\alpha_{1}+\alpha_{2}=1$ with $\alpha_{1}\ge 0$ and $\alpha_{2}\ge 0$. In the above equation, $\alpha_{1}$ and $\alpha_{2}$ depends mainly on the following factors:  $\hat{\rho}_{TDS}, \hat{\rho}_{RTS}, N_{TDS}$, and  $N_{RTS}$, where $\hat{\rho}_{TDS}$, and $\hat{\rho}_{RTS}$ are the receive  SNRs at Alice in the training and real-time measuring stages, respectively. Based on this, we propose three convex weight combinators with weight factors as follows
\begin{align}\label{SNR-Weighting}
\alpha_1=\frac{\hat{\rho}_{TDS}}{\hat{\rho}_{TDS}+\hat{\rho}_{RTS}},\nonumber\\
\alpha_2=\frac{\hat{\rho}_{RTS}}{\hat{\rho}_{TDS}+\hat{\rho}_{RTS}},
\end{align}
\begin{align}\label{Times-Weighting}
\alpha_1=\frac{N_{TDS}}{N_{TDS}+N_{RTS}},\nonumber\\
\alpha_2=\frac{N_{RTS}}{N_{TDS}+N_{RTS}},
\end{align}
and
\begin{align}\label{}
\alpha_1=\frac{\hat{\rho}_{TDS}\cdot N_{TDS}}{\hat{\rho}_{TDS}\cdot N_{TDS}+\hat{\rho}_{RTS}\cdot N_{RTS}},\nonumber\\
\alpha_2=\frac{\hat{\rho}_{RTS}\cdot N_{RTS}}{\hat{\rho}_{TDS}\cdot N_{TDS}+\hat{\rho}_{RTS}\cdot N_{RTS}},
\end{align}
respectively. Here, for the sake of convenience, in what follows,  (\ref{SNR-Weighting}), (\ref{Times-Weighting}) and (\ref{Both-Weighting}) are called the proposed weight methods \Rmnum{1},~\Rmnum{2},~ and  \Rmnum{3}, respectively. Similarly, we have the corresponding weight combiners for DOA variance estimation
\begin{align}
\hat{\sigma^2}_{ML}&=\alpha_{1}\tilde{\sigma}^2_{TDS}+\alpha_{2}\hat\sigma^2_{NDS}
\end{align}
where the weight parameters $\alpha_{1}$ and $\alpha_{2}$ are given by (\ref{SNR-Weighting}),~(\ref{Times-Weighting}),~ and  (\ref{Both-Weighting}), respectively. And
\begin{equation}
\tilde{\sigma}^2_{TDS}=\frac{\hat{\rho}_{TDS}}{\hat{\rho}_{RTS}}\hat\sigma^2_{TDS}
\end{equation}

In the moving scenario, it is particularly noted  that the  learned variance in training stage has an intimately direct relationship with real-time measuring    one. At least it is used to predict the variance of the real-time measuring case in terms of their SNRs, and numbers of RTS and TDS. However, due to moving, the direction angles of Bob and Eve will change with time, the learned DOA mean in RTS will be independent of the learned DOA mean from TDS. In other words, the learned DOA mean from  RTS will be used as the final learned DOA. In such a situation, the weight factors $\alpha_{1}$ and  $\alpha_{2}$ in (\ref{ML-DOA-ConWei}) are set to 1 and 0, respectively.

\section{Proposed robust HAD beamformer for DM}
 In the above section, we have attained  the measured value $\hat\theta_{ML}$ and variance $\hat{\sigma^2}_{ML}$ of DOA, then we have  the ML-based DOA output modeled as
  \begin{align}\label{DOA-Error-Mod}
  \theta=\hat{\theta}_{ML}+\Delta\hat{\theta}
  \end{align}
with its PDF as follows
  \begin{align}\label{DOA-PDF2}
f\left(\theta;\hat{\theta}_{ML},\hat{\sigma}^2_{ML}\right)=\frac{1}{\sqrt{2\pi\hat{\sigma}^2_{ML}}}\exp\left\{-\frac{1}{2\hat{\sigma}^2_{ML}}\left(\theta-\hat{\theta}_{ML}\right)^2\right\},
\end{align}
which yields the PDF of DOAME
  \begin{align}\label{DOA-PDF1}
f\left(\Delta\hat{\theta}; \hat{\sigma}^2_{ML}\right)=\frac{1}{\sqrt{2\pi\hat{\sigma}^2_{ML}}}\exp\left\{-\frac{1}{2\hat{\sigma}^2_{ML}}\left(\Delta\hat{\theta}\right)^2\right\}.
\end{align}
Considering an important fact that the DOAME falls into the interval $[\pi,\pi]$ , similar to \cite{Wu, Zhou}, the PDF of  DOAME can be approximated by the truncated Gaussian density function
\begin{align}
f(\Delta\hat{\theta})=
\begin{cases}
\frac{1}{K_d\sqrt{2\pi\hat{\sigma}^2_{ML}}}\exp\left(-\frac{(\Delta\hat{\theta})^2}{2\hat{\sigma}^2_{ML}}\right),&{\frac{\Delta\hat{\theta}}{\Delta\theta_{\max}} \in[-1,1]}\\
0,\ &\mbox{others}
\end{cases}
\end{align}
where the truncated factor $K_d$ is given by
\begin{align}
K_d=\int_{-\Delta\theta_{\max}}^{\Delta\theta_{\max}}\frac{1}{\sqrt{2\pi\hat{\sigma}^2_{ML}}}\exp\left(-\frac{({\Delta\hat\theta})^2}{2\hat{\sigma}^2_{ML}}\right)d(\Delta\hat\theta),
\end{align}
and $\Delta\theta_{\max}$ may be chosen to be a value less than or equal to $\pi$ but larger than a multiple of $\sigma$. For example, $\Delta\theta_{\max}$ can be taken to be half of the main-beam width.
\begin{figure}[h]
  \centering
  \includegraphics[width=9cm]{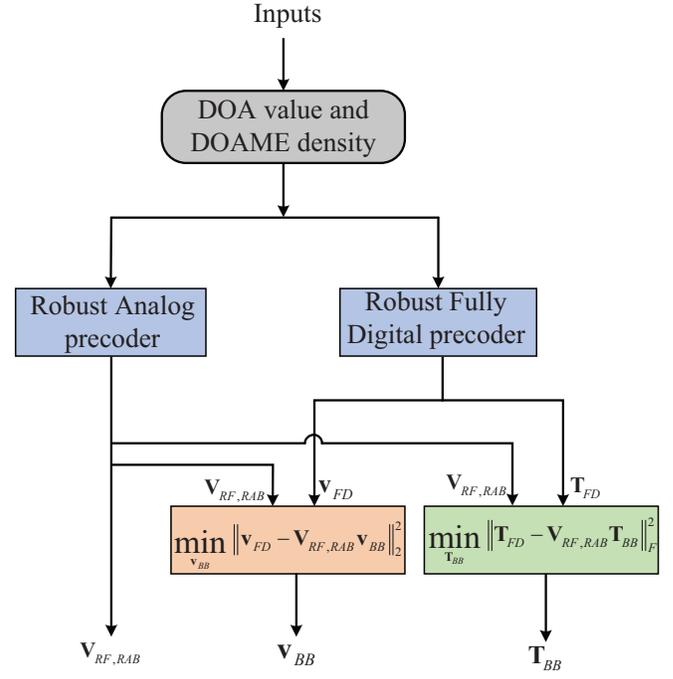}\\
  \caption{Block diagram of the proposed  robust DM  synthesis method}
  \label{HAD_DM}
\end{figure}
 In accordance with the above approximation and in terms of conditional expectation,  three robust beammforming matrices $\mathbf{V}_{RF}$, $\mathbf{v}_{BB}$, and $\mathbf{T}_{BB}$,  are in order presented for HAD-based DM systems in the following.

\subsection{Proposed robust Analog beamforming matrix $\mathbf{V}_{RF}$}

Since we have already obtained the direction of desired user in Section III, a direct method to design the analog precoder is to directly implement  phase alignment (PA), which aligns the phases of phase shifters of subarray $k$ in analog part to the direction of desired user. The block submatrix $k$ $\mathbf{v}_k$ in $\mathbf{V}_{RF}$  is given by
\begin{align}
\mathbf{v}_{NRAB,k}\left(\hat{\theta}_{ML}\right)=\mathbf{h}_k\left(\hat{\theta}_{ML}\right)
\end{align}
where the SV  $\mathbf{h}_k(\bullet)$  has been defined in  (\ref{Subarray-k-SV}). The PA method to design the analog precoder in (46) doesn't take the DOA estimation error into consideration, which  completely depends on the precision of DOA estimation. However, there  always exists measurement error in the measured angle as shown in (\ref{ML-DOA-ConWei}). To combat this type of errors, we will exploit the statistical property of DOAME, i.e. its PDF to propose a robust analog beamformer (RAB). In this precoder, given the PDF of  DOAME and measured DOA, minimizing the  mean square  error criterion directly yields the following robust beamforming vector for subarray $k$
\begin{align}
\mathbf{v}_{RAB,k}&=\text{E}\left\{\mathbf{h}_k\left(\theta\right)|\hat{\theta}_{ML}, f\left(\Delta\theta_{ML}\right)\right\}\nonumber\\
&=\text{E}_{\Delta\theta_{ML}}\left\{\mathbf{h}_k\left(\hat{\theta}_{ML}+\Delta\theta_{ML}\right)\right\}
\end{align}
where the $m$ th element of $\mathbf{v}_{RAB,k}$ is derived in Appendix A. Placing all $\mathbf{v}_{RAB,k}$ into  (\ref{V-RF}), we obtain the robust analog beamforming matrix $\mathbf{{V}}_{RF, RAB}$.

\subsection{Proposed robust digital beamforming vector $\mathbf{v}_{BB}$}
Now, we compute the robust digital beamforming (RDB) vector $\mathbf{v}_{BB}$. Similar to \cite{Xianghao, Heath2016An}, the problem of solving $\mathbf{v}_{BB}$ can be casted as
\begin{align}\label{Opt-V-BB}
&\min\limits_{\mathbf{v}_D}{\kern 1pt}~~~~~~\parallel\mathbf{v}_{FD}-\mathbf{V}_{RF,RAB}\mathbf{v}_{BB}\parallel_2^2 \nonumber\\
&\text{subject~to}~~~~~\parallel\mathbf{V}_{RF,RAB}\mathbf{v}_{BB}\parallel^2=1.
\end{align}
where $\mathbf{v}_{FD}$ stands for the optimal FD beamforming vector of confidential message. Observing the above the above optimization problem, we find an obvious fact: if we design  $\mathbf{v}_{FD}$ well, then we easily address the problem of optimzing  $\mathbf{v}_{BB}$ with $\mathbf{V}_{RF,RAB}$ available in the precedeing subsection.   Here, $\mathbf{v}_{FD}$  is chosen to be the beamforming vector of confidential message of robust null-space projection method in \cite{Siming}.
Due to the special structure of $\mathbf{V}_{RF,RAB}$, we have the identity $\mathbf{V}^H_{RF,RAB}\mathbf{V}_{RF,RAB}=\mathbf{I}$, and further $\parallel\mathbf{V}_{RF,RAB}\mathbf{v}_{BB}\parallel^2=\parallel\mathbf{v}_{BB}\parallel^2=1$.
So the optimization problem in (\ref{Opt-V-BB}) can be simplified as
\begin{align}\label{Opt-V-BB-Simp}
&\min\limits_{\mathbf{v}_{BB}}{\kern 1pt}~~~~~~~~~~\parallel\mathbf{v}_{FD}-\mathbf{V}_{RF,RAB}\mathbf{v}_{BB}\parallel^2_2 \nonumber\\
&\text{subject~to}~~~~~\parallel\mathbf{v}_{BB}\parallel^2=1
\end{align}
whose objective function can be expanded as
\begin{align}
&g\left(\mathbf{v}_{BB}\right)=\parallel\mathbf{v}_{FD}-\mathbf{V}_{RF,RAB}\mathbf{v}_{BB}\parallel^2_2 \nonumber\\
&=(\mathbf{v}_{FD}-\mathbf{V}_{RF,RAB}\mathbf{v}_{BB})^H(\mathbf{v}_{FD}-\mathbf{V}_{RF,RAB}\mathbf{v}_{BB})\nonumber\\
&=\mathbf{v}_{FD}^H\mathbf{v}_{FD}-\mathbf{v}_{FD}^H\mathbf{V}_{RF,RAB}\mathbf{v}_{BB}-\mathbf{v}_{BB}^H\mathbf{V}_{RF,RAB}^H\mathbf{v}_{FD}\nonumber\\
&~~~~+\mathbf{v}_{BB}^H\mathbf{V}_{RF,RAB}^H\mathbf{V}_{RF,RAB}\mathbf{v}_{BB}\nonumber\\
&=\mathbf{v}_{FD}^H\mathbf{v}_{FD}-\mathbf{v}_{FD}^H\mathbf{V}_{RF,RAB}\mathbf{v}_{BB}\nonumber\\
&-\mathbf{v}_{BB}^H\mathbf{V}_{RF,RAB}^H\mathbf{v}_{FD}+\mathbf{v}_{BB}^H\mathbf{v}_{BB}.
\end{align}
Differentiating the objective function $g\left(\mathbf{v}_{BB}\right)$ with respect to $\mathbf{v}_{BB}$ and setting it equal to zero yields
\begin{align}
\frac{\partial g\left(\mathbf{v}_{BB}\right)}{\partial\mathbf{v}_{BB}^H}
=\mathbf{v}_{BB}-\mathbf{V}_{RF,RAB}^H\mathbf{v}_{FD}=0,
\end{align}
which gives
\begin{align}
\mathbf{v}_{BB}=\mathbf{V}_{RF,RAB}^H\mathbf{v}_{FD},
\end{align}
which is normalized to a unit vector
\begin{align}
\mathbf{v}_{RBB,~BB}=\frac{\mathbf{V}_{RF,RAB}^H\mathbf{v}_{FD}}{\parallel\mathbf{V}_{RF,RAB}^H\mathbf{v}_{FD}\parallel_2}.
\end{align}
with the constraint $\mathbf{v}_{BB}^H\mathbf{v}_{BB}$=1 in (\ref{Opt-V-BB-Simp}). This completes the construction of $\mathbf{v}_{BB}$.

\subsection{Proposed robust AN projection matrix $\mathbf{T}_{BB}$}
In this subsection, we will optimize the design of the digital AN projection matrix $\mathbf{T}_{BB}$. Actually, its designing is very similar that of $\mathbf{v}_{BB}$ in the previous subsection. In the same fashion,
after the FD AN projection matrix $\mathbf{T}_{BB}$ is set to the robust NSP matrix in \cite{Siming},  the optimization problem of finding $\mathbf{T}_{BB}$  can be casted as
\begin{align}
&\min\limits_{\mathbf{T}_{BB}}{\kern 1pt}~~~~~~~~~~\parallel\mathbf{T}_{FD}-\mathbf{V}_{RF,RAB}\mathbf{T}_{BB}\parallel^2_F \nonumber\\
&\text{subject~to}~~~~~\parallel\mathbf{V}_{RF,RAB}\mathbf{T}_{BB}\parallel^2_F=1\nonumber\\
\end{align}
Because of $\parallel\mathbf{V}_{RF,RAB}\mathbf{T}_{BB}\parallel_F^2=\parallel\mathbf{T}_{BB}\parallel_F^2=1$, we can further simplify the above optimization problem  as
\begin{align}
&\min\limits_{\mathbf{T}_{BB}}{\kern 1pt}~~~~~~~~~~\parallel\mathbf{T}_{FD}-\mathbf{V}_{RF,RAB}\mathbf{T}_{BB}\parallel^2_F \nonumber\\
&\text{subject~to}~~~~~\parallel\mathbf{T}_{BB}\parallel^2_F=1\nonumber\\
\end{align}
The objective function $\parallel\mathbf{T}_{opt}-\mathbf{V}_{RF,RAB}\mathbf{T}_{BB}\parallel^2_F$ can be represented as
\begin{align}
&\parallel\mathbf{T}_{FD}-\mathbf{V}_{RF,RAB}\mathbf{T}_{BB}\parallel^2_F\nonumber\\
&=\parallel\text vec(\mathbf{T}_{FD}-\mathbf{V}_{RF,RAB}\mathbf{T}_{BB})\parallel_2^2\nonumber\\
&=\parallel\text vec(\mathbf{T}_{FD})-\text vec(\mathbf{V}_{RF,RAB}\mathbf{T}_{BB})\parallel_2^2\nonumber\\
&=\parallel\text vec(\mathbf{T}_{FD})-(\mathbf{I}_K\otimes\mathbf{V}_{RF,RAB})\text vec(\mathbf{T}_{BB})\parallel_2^2\nonumber\\
&=\parallel\mathbf{g}-\mathbf{Q}\mathbf{t}\parallel_2^2
\end{align}
where $\mathbf{g}=\text vec(\mathbf{T}_{FD})$, $\mathbf{Q}=\mathbf{I}_K\otimes\mathbf{V}_{RF,RAB}$ and $\mathbf{t}=\text{vec}(\mathbf{T}_{BB})$. And also the constraint is reformulated as $\parallel\mathbf{t}\parallel_2^2=1$.
Hence, the original optimization problem in (42) reduces to
\begin{align}
&\min\limits_{\mathbf{T}_D}{\kern 1pt}~~~~~~~~~~\parallel\mathbf{g}-\mathbf{Q}\mathbf{t}\parallel^2_2 \nonumber\\
&\text{subject~to}~~~~~\parallel\mathbf{t}\parallel^2_2=1,
\end{align}
whose the objective function can be expanded as
\begin{align}
&\parallel\mathbf{g}-\mathbf{Q}\mathbf{t}\parallel^2_2 \nonumber\\
&=(\mathbf{g}-\mathbf{Q}\mathbf{t})^H(\mathbf{g}-\mathbf{Q}\mathbf{t})\nonumber\\
&=\mathbf{g}^H\mathbf{g}-\mathbf{g}^H\mathbf{Q}\mathbf{t}-\mathbf{t}^H\mathbf{Q}^H\mathbf{g}+\mathbf{t}^H\mathbf{Q}^H\mathbf{Q}\mathbf{t}\nonumber\\
&\mathop {= }\limits^{(a)}\mathbf{g}^H\mathbf{g}-\mathbf{g}^H\mathbf{Q}\mathbf{t}-\mathbf{t}^H\mathbf{Q}^H\mathbf{g}+\mathbf{t}^H\mathbf{Q}^H\mathbf{Q}\mathbf{t}.
\end{align}
Because
\begin{align}
\mathbf{t}^H\mathbf{Q}^H\mathbf{Q}\mathbf{t}&=\mathbf{t}^H(\mathbf{I}_K\otimes\mathbf{V}_{RF})^H(\mathbf{I}_K\otimes\mathbf{V}_{RF})\mathbf{t}\nonumber\\
&=\mathbf{t}^H(\mathbf{I}_K^H\otimes\mathbf{V}_{RF}^H)(\mathbf{I}_K\otimes\mathbf{V}_{RF})\mathbf{t}\nonumber\\
&=\mathbf{t}^H(\mathbf{I}_K^H\mathbf{I}_K\otimes\mathbf{V}_{RF}^H\mathbf{V}_{RF})\mathbf{t}\nonumber\\
&=\mathbf{t}^H(\mathbf{I}_K\otimes\mathbf{I}_K)\mathbf{t}\nonumber\\
&=\mathbf{t}^H\mathbf{t},
\end{align}
the objective function reduces to
\begin{align}
&\parallel\mathbf{g}-\mathbf{Q}\mathbf{t}\parallel^2_2=\mathbf{g}^H\mathbf{g}-\mathbf{g}^H\mathbf{Q}\mathbf{t}-\mathbf{t}^H\mathbf{Q}^H\mathbf{g}+\mathbf{t}^H\mathbf{t}.
\end{align}
Taking the derivative of $\parallel\mathbf{g}-\mathbf{Q}\mathbf{t}\parallel^2_2$ with respect to $\mathbf{t}^H$ and setting it to zero, we have
\begin{align}
\frac{\partial \parallel\mathbf{g}-\mathbf{Q}\mathbf{t}\parallel^2_2}{\partial\mathbf{t}^H}
=\mathbf{t}-\mathbf{Q}^H\mathbf{g}=0,
\end{align}
Similarly, the optimal solution is
\begin{align}\label{t-opt}
\mathbf{t}_{opt}=\frac{\mathbf{Q}^H\mathbf{g}}{\parallel\mathbf{Q}^H\mathbf{g}\parallel}.
\end{align}
where $\mathbf{Q}^H\mathbf{g}$ can be rewritten in the following form
\begin{align}
\mathbf{Q}^H\mathbf{g}=&\left[ {\begin{array}{*{20}{c}}
{\mathbf{V}_{RF,RAB}^H}&{\mathbf{0}}&{\cdots}&{\mathbf{0}}\\
{\mathbf{0}}&{\mathbf{V}_{RF,RAB}^H}&{\cdots}&{\mathbf{0}}\\
{\vdots}&{\vdots}&{\ddots}&{\vdots}\\
{\mathbf{0}}&{\mathbf{0}}&{\cdots}&{\mathbf{V}_{RF,RAB}^H}
\end{array}} \right]\left[ {\begin{array}{*{20}{c}}
{\mathbf{g}_1}\\
{\mathbf{g}_2}\\
{\vdots}\\
{\mathbf{g}_N}
\end{array}} \right]\nonumber\\
=&\left[ {\begin{array}{*{20}{c}}
{\mathbf{V}_{RF,RAB}^H\mathbf{g}_1}\\
{\mathbf{V}_{RF,RAB}^H\mathbf{g}_2}\\
{\vdots}\\
{\mathbf{V}_{RF,RAB}^H\mathbf{g}_N}
\end{array}} \right].
\end{align}
Using (\ref{t-opt}) and $\mathbf{t}=\text{vec}(\mathbf{T}_{BB})$, the robust digital AN projection matrix $\mathbf{T}_{opt}$ is readily gotten by just the reverse operation of matrix-to-vector operator $\text{vec}(\bullet)$.

\section{Simulation Results and discussions}
In this section, simulation results are presented to evaluate the performance of the proposed I-HAD-ESPRIT DOA estimator,its enhanced version ML-based scheme,  and robust HAD-based beamforming schemes for DM. Simulation parameters are chosen as follows:  $\theta_d=50^\circ$ and $\theta_e=70^\circ$.  The number of snapshots per DOA measurement is set to be $L=64$,  and the total number of antennas at Alice is $N=64$ with $M \in\{4,~8,~16$\}. The PA factor $\beta=0.9$ and the antenna spacing is $d=\frac{\lambda}{2}$. $\sigma_d^2=\sigma_e^2=\sigma^2$ and QPSK modulation is used to assess bit error rate (BER) performance.

\begin{figure}[h]
  \centering
  \includegraphics[width=9cm,height=8cm]{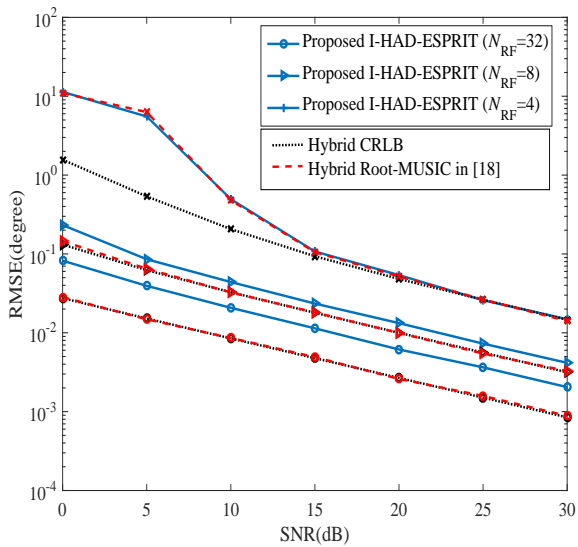}\\
  \caption{RMSE versus SNR of the proposed ESPRIT with $N=32 with M=\{4,8,16\}$.}
  \label{DOA-HAD-ESPRT}
\end{figure}

\begin{figure}[h]
  \centering
  \includegraphics[width=9cm,height=8cm]{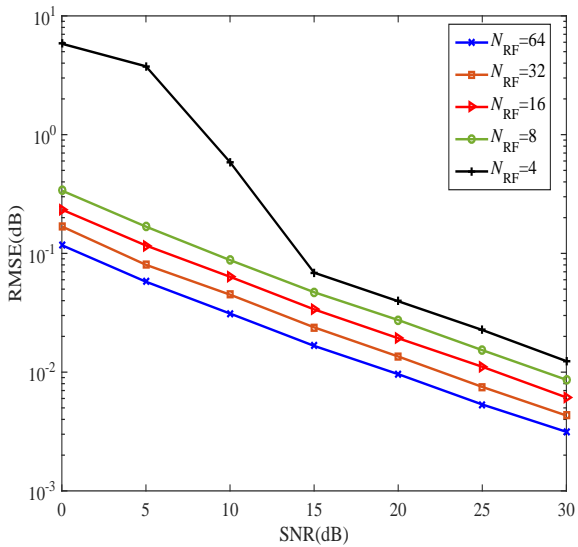}\\
  \caption{RMSE versus SNR of the proposed SNR estimator.}
  \label{SNR-SNR}
\end{figure}

\begin{figure}
  \centering
  \includegraphics[width=9cm,height=8cm]{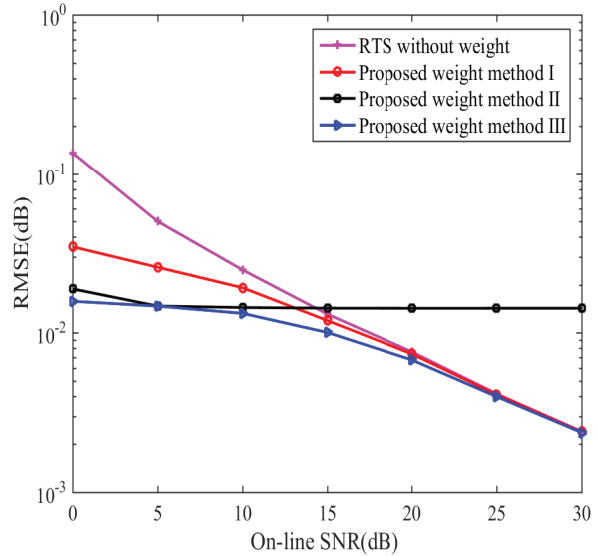}\\
  \caption{RMSE versus of SNR of the proposed three  weight combiners of DOA.}
 \label{angle-weight}
\end{figure}

\begin{figure}[h]
\centering
\includegraphics[width=9cm,height=8cm]{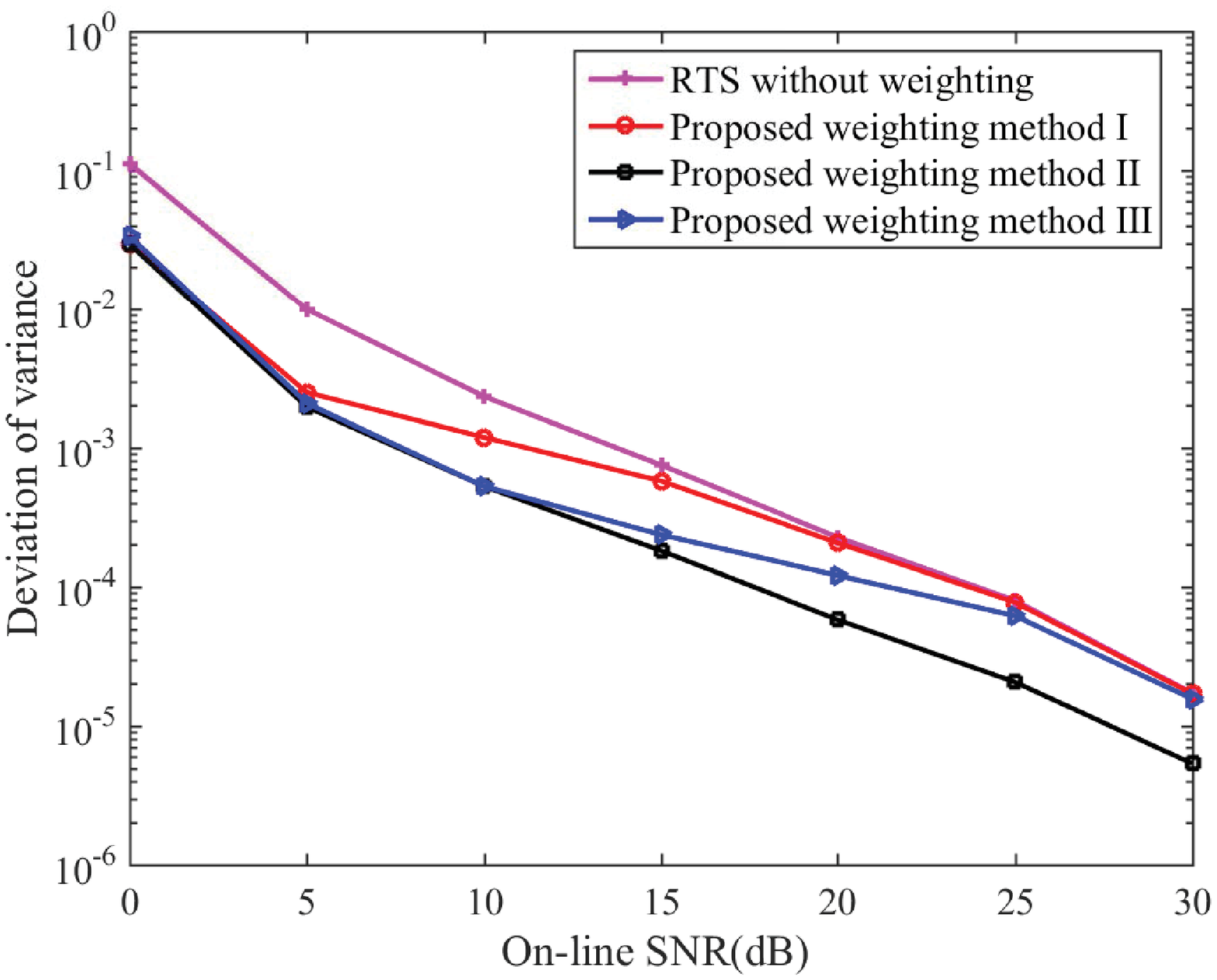}
\caption{RMSE versus SNR of three weight methods of estimating variance ($\mathrm{SNR_{TDS}=10dB}$).}
\label{var-weight}
\end{figure}

\begin{figure}[h]
\centering

\includegraphics[width=9cm,height=8cm]{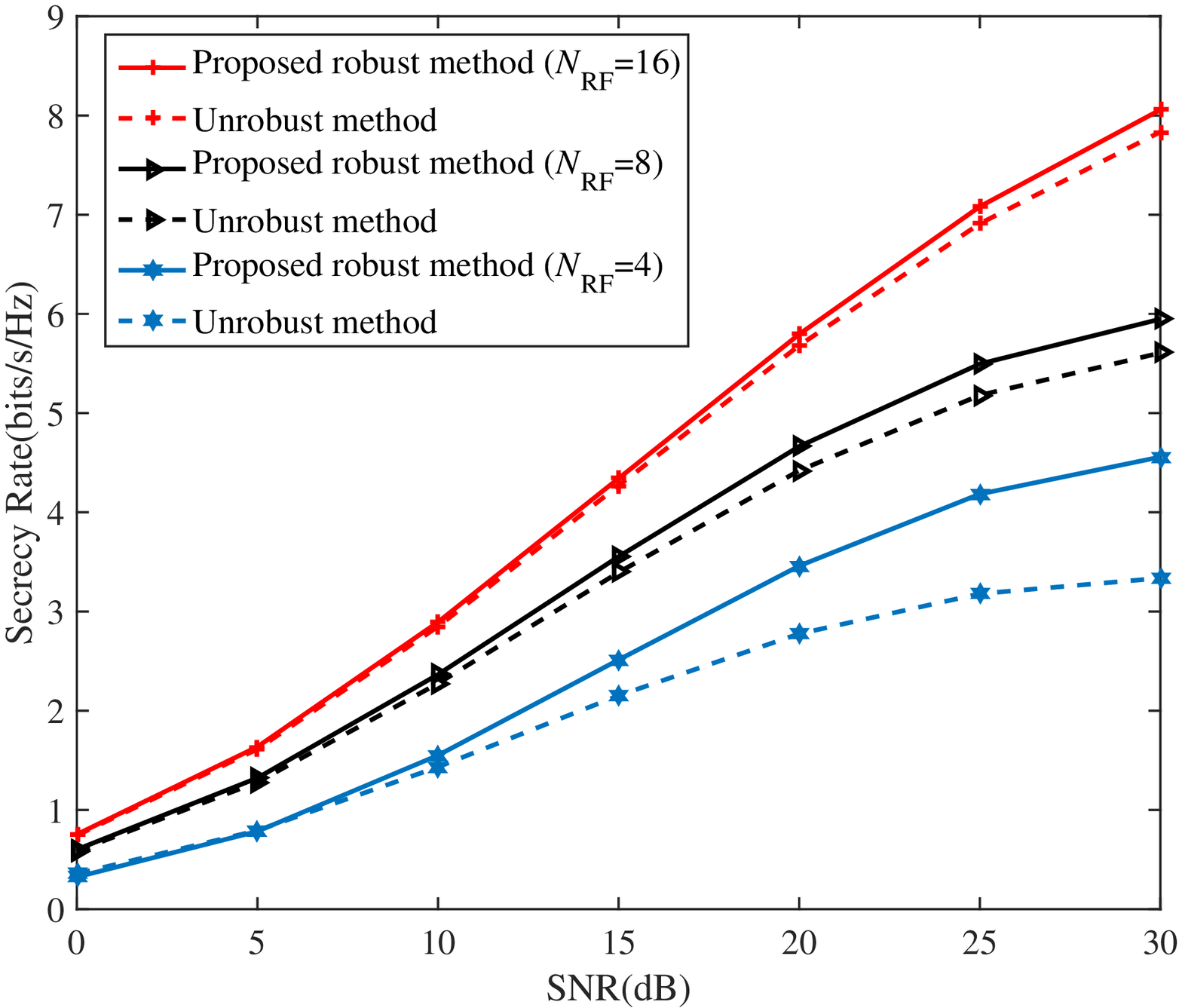}
\caption{Curves of SR versus SNR of the proposed robust synthesis method for HAD-based DM with $SNR_{TDS}=-5$dB.}
\label{Cur-Rob-SR}
\end{figure}

\begin{figure}[h]
\centering
\includegraphics[width=9cm,height=8cm]{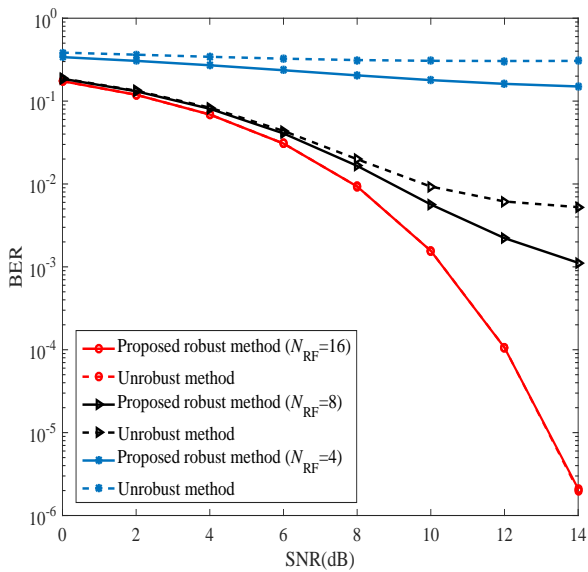}
\caption{Curves of BER versus SNR of the proposed robust synthesis method for HAD-based DM with $SNR_{TDS}=-5$dB.}
\label{Cur-Rob-BER}
\end{figure}

Fig.~\ref{DOA-HAD-ESPRT} demonstrates the performance of root mean squared error(RMSE) versus SNR of the proposed I-HAD-ESPRIT method with $N=64, L=64$ and $M\in\{2,8,16\}$. As can be seen from Fig.~\ref{DOA-HAD-ESPRT}, we find an interesting phenomenon that the performance gap between I-HAD-ESPRIT and hybrid CRLB  becomes smaller and smaller as the value of $M$ increases from 2 to 16. Eventally,  this gap tends to zero. And when $M$ is up to 16 (large enough), the proposed I-HAD-ESPRIT can achieve has the same performance as the hybrid Root-MUSIC in \cite{Qin} and achieves hybrid CRLB in the medium and high SNR regions. However, the proposed I-HAD-ESPRIT performs slightly worse than  hybrid Root-MUSIC in \cite{Qin} for most cases. The main reason is as follows: when $M$ increases, the number of RF chains will accordingly reduce such that a smaller total number of spatial-time sampling points are harvested to calculate covariance matrix of receive signal.

Fig.~\ref{SNR-SNR} plots the RMSE performance of the estimated SNR versus exact SNR.  Similar to Fig.~\ref{DOA-HAD-ESPRT}, the performance curves of SNR measurement  shows the same trend as that in Fig.~\ref{DOA-HAD-ESPRT}. As the number of subarray antennas increases, the  precision of SNR estimation decreases gradually. In other words,  reducing the number of RF chains will result in a RMSE performance loss over the estimated SNR.

Fig.~\ref{angle-weight} demonstrates the curves of variances of the proposed three different weight DOA combiners   versus SNR for SNR$_{TDS}$=5dB. The learning output of RTS is used as a performance upper bound. Due to the fact that the RTS has fewer number of values of DOA measurement compared to TDS, it performs worst. It can be seen from Fig.~\ref{angle-weight}:  the three proposed weight methods performs better than  the learning output of RTS without use of TDS in the low SNR region because they all fully exploit the DTS to improve the estimate precision.  The proposed method III is the best one among the three proposed methods due to the fact it not only utilizes the element numbers of TDS and  RTS but also exploits the SNR difference of TDS and RTS.

Fig.~\ref{var-weight} illustrates the curves  of  variances of the proposed three weight  estimators of angle variance versus SNR with SNR$_{TDS}$=10dB. The learning output of RTS  is adopted as a performance benchmark. It can be seen from Fig.~\ref{var-weight} that the proposed three  weight methods exceeds the learning output of RTS without using TDS in terms of variance performance.  Due to the fact that the learning output of TDS is normalized to the same magnitude of the estimation output of RTS, the element numbers of the two sets TDS and RTS will play a dominant role in the weighting process. The result can be seen from  Fig.~\ref{var-weight}. That is, the proposed method II is the best one among the proposed three methods.

Fig.~\ref{Cur-Rob-SR} plots the curves SR versus SNR of  our proposed robust synthesis method for  DM using HAD structure with SNR$_{TDS}$=-5dB . Here,  the corresponding non-robust method is used as a performance benchmark and the number of RF chains is chosen as  follows: $N_{RF}=\{4,~8,~16\}$. As can be seen from Fig.~\ref{Cur-Rob-SR}, when  DOA measurement error exists, the proposed robust DM method always performs better than the corresponding non-robust one especially in the medium and high SNR regions. In particular, with decrease in the number of RF chains, the performance gain of the proposed robust method over nonrobust one become more obvious. In such a situation, reducing the number of RF chains, under the condition that other parameters are  fixed like  the receive SNR and a total number of antennas at Alice,  the DOA measurement error become large due to a smaller number of spatial-time sampling points available. In other words,  the performance gain achieved by exploiting the density distribution of DOA measurement error becomes more obvious in such a scenario. Otherwise, as the number of RF chains increases,  a larger number of spatial-time sampling points is available to compute a high-precision variance matrix in ESPRIT. A higher-precision DOA estimation is gotten. This implies a smaller measurement error, which will accordingly decrease the performance gain achieved by the robust one proposed  by us over nonrobust one.

 Fig.~\ref{Cur-Rob-BER} shows the BER performance versus SNR  of our proposed robust hybrid DM with the corresponding  non-robust one as a performance reference. Here,  $N=64$, and $M=\in\{4,~8,~16\}$. As shown in Fig.~\ref{Cur-Rob-BER}, we can find that the proposed robust method achieves a better BER performance compared with the corresponding non-robust one for $N_{RF}=\in\{4,~8,~16\}$.  As the number $N_{RF}$ of RF chains increases, the BER performance difference between robust and non-robust ones disappears.  The performance tendency in this figure is very similar to that in Fig.~\ref{Cur-Rob-SR}.

\section{Conclusion}
In this paper, an I-HAD-ESPRIT DOA measurement method was proposed for HAD receiver. Here, the phase ambiguity from HAD structure  is addressed successfully. The proposed I-HAD-ESPRIT can achieve the HAD CRLB. In the mean time, we also proposed a SNR estimator. By using the proposed I-HAD-ESPRIT method and histogram method in ML, we find an important fact that the measured DOA or DOAME follows a Gaussian distribution. Subsequently, by applying the proposed I-HAD-ESPRIT method repeatedly, we established two data sets: TDS and RTS. After the maximum likelihood learning method was optimized over TDS and RTS, we obtained the two high-precision learning  outputs of DOA and variance. To combine the two learning outputs and improve performance, three weight methods was proposed to exploit the SNRs, numbers of TDS and RTS, or both. For DOA, the proposed weight method III of exploiting both is the best one among the three methods whereas the proposed weight method II is the best one among the three methods. Finally, based on the above, a robust  synthesis method for HAD-based DM transmitter was proposed to exploit the density distribution of DOA measurement error, with emphasis on the design of robust analog beamforming vector. Compared to the corresponding non-robust method, the proposed robust method can achieve a substantial SR performance gain.


.\section*{Appendix A:~Derivation of Estimators of DOA mean and variance }
From Section III,  the measured value $\tilde{\theta}$ of DOA is shown to be a Gaussian random variable with mean $\theta$ and variance $\sigma^2$, we have its PDF as follows
\begin{align}\label{3}
f\left(\tilde\theta(n);\theta,\sigma^2\right)=\frac{1}{\sqrt{2\pi\sigma^2}}\exp\{-\frac{1}{2\sigma^2}(\tilde\theta(n)-\theta)^2\}
\end{align}
Given $N_{DOA}$  measured values of angle of Bob: $\left(\tilde{\theta}_1,\cdots,\tilde{\theta}_{N_{DOA}}\right)$,  we have their likelihood
\begin{align}\label{4}
&f\left(\tilde{\theta}_1,\cdots,\tilde{\theta}_{N_{DOA}};\theta,\sigma^2\right)\nonumber\\
&=(\frac{1}{2\pi\sigma^2})^\frac{N}{2}\exp\{-\frac{1}{2\sigma^2}\sum_{i=0}^{N-1}(\tilde\theta(n)-\theta)^2\},
\end{align}
which gives the corresponding log-likelihood function
\begin{align}
&\ln f\left(\tilde{\theta}_1,\cdots,\tilde{\theta}_{N_{DOA}};\theta,\sigma^2\right)\nonumber\\
&=-\frac{N_{DOA}}{2}\ln (2\pi\sigma^2)-\frac{1}{2\sigma^2}\sum_{i=0}^{N_{DOA}-1}(\tilde{\theta}_i-\theta)^2.
\end{align}
Then, we compute the first derivative of $\ln P(\mathbf{\theta_r}|\theta)$ with respect to $\theta$ as follows
\begin{align}
\frac{\partial \ln  f\left(\tilde{\theta}_1,\cdots,\tilde{\theta}_{N_{DOA}};\theta,\sigma^2\right)}{\partial\theta}=\frac{1}{\sigma^2}\sum_{i=0}^{N_{DOA}-1}(\tilde{\theta}_i-\theta)
\end{align}
which is set to  zero, then we can obtain the estimated value of $\theta$
\begin{align}
\hat{\theta}=\frac{1}{N_{DOA}}\sum_{i=0}^{N_{DOA}}\tilde{\theta}_i
\end{align}
In the same way, we set the first derivative of $\ln P(\mathbf{\theta_r}|\theta)$ with respect to $\sigma^2$ to zero
\begin{align}
&\frac{\partial \ln  f\left(\tilde{\theta}_1,\cdots,\tilde{\theta}_{N_{DOA}};\theta,\sigma^2\right)}{\partial\sigma^2}\nonumber\\
&=-\frac{N_{DOA}-1}{2\sigma^2}+\frac{1}{2\sigma^4}\sum_{i=0}^{N_{DOA}-1}(\tilde{\theta}_i-\theta)^2=0,
\end{align}
which yields
\begin{align}
\hat{\sigma^2}=\frac{1}{N_{DOA}}\sum_{i=1}^{N_{DOA}}(\tilde{\theta}_i-\theta)^2.
\end{align}
Since the true value of $\theta$ is unknown, replacing $\theta$ with its estimated value $\hat{\theta}$  yields the bootstrap solution
\begin{align}\label{Var-Est-1}
\hat{\sigma^2}=\frac{1}{N_{DOA}}\sum_{i=1}^{N_{DOA}}(\tilde{\theta}_i-\hat\theta)^2
\end{align}
in accordance with \cite{Larry}.
 However, the above solution is a biased estimator.  Instead, we prefer the following unbiased estimator
 \begin{align}\label{Var-Est-2}
\hat{\sigma^2}=\frac{1}{N_{DOA}-1}\sum_{i=1}^{N_{DOA}}(\tilde{\theta}_i-\hat\theta)^2
\end{align}
with the same form as that of (\ref{Var-Est-1}).

.\section*{Appendix B:~Derivation of RAB $\mathbf{\tilde{V}_{RF}}$}

Considering the DOA measurement error, the nonzero element of $\mathbf{V}_{RF,RAB}$ can be represented as
\begin{align}
&\hat{v}_{k,m}\nonumber\\
&=\int_{-\Delta\theta_\mathrm{{max}}}^{\Delta\theta_\mathrm{{max}}}e^{-j\frac{2\pi d}{\lambda}[(k-1)M+m-\frac{N+1}{2}]\cos(\hat{\theta}-\Delta\theta)}\cdot p(\Delta\theta)d(\Delta\theta)\nonumber
\end{align}
\begin{align}
&=\int_{-\Delta\theta_\mathrm{{max}}}^{\Delta\theta_\mathrm{{max}}}e^{-j\alpha_{k,m}\cos(\hat{\theta}-\Delta\theta)}\cdot p(\Delta\theta)d(\Delta\theta)\nonumber\\
&=\int_{-\Delta\theta_\mathrm{{max}}}^{\Delta\theta_\mathrm{{max}}}e^{-j\alpha_{k,m}[\cos(\hat{\theta})\cos(\Delta\theta)+\sin(\hat{\theta})\sin(\Delta\theta)]}\cdot p(\Delta\theta)d(\Delta\theta)\nonumber\\
\label{69}
\end{align}
where
\begin{align}
\alpha_{k,m}=\frac{2\pi d}{\lambda}[(k-1)M+m-\frac{N+1}{2}].
\end{align}
 By utilizing the second-order Taylor expansion, we can expand $\cos(\Delta\theta)$ and $\sin(\Delta\theta)$ at point $\Delta\theta=0$ as
\begin{align}\label{Taylor-exp}
&\cos(\Delta\theta)\approx1-\frac{1}{2}(\Delta\theta)^2\nonumber\\
&\sin(\Delta\theta)\approx\Delta\theta.
\end{align}
Substituting (\ref{Taylor-exp}) in  (\ref{69}) yields
\begin{align}\label{Ing-Analog}
&\hat{v}_{k,m}\nonumber\\
&=\int_{-\Delta\theta_\mathrm{{max}}}^{\Delta\theta_\mathrm{{max}}}e^{-j\alpha_{k,m}[\cos(\hat{\theta})\cos(\Delta\theta)+\sin(\hat{\theta})\sin(\Delta\theta)]}\cdot p(\Delta\theta)d(\Delta\theta)\nonumber\\
&=\int_{-\Delta\theta_\mathrm{{max}}}^{\Delta\theta_\mathrm{{max}}}e^{-j\alpha_{k,m}[\cos(\hat{\theta})-\cos(\hat{\theta})\cdot\frac{1}{2}(\Delta\theta)^2+\sin(\hat{\theta})(\Delta\theta)]} p(\Delta\theta)d(\Delta\theta)\nonumber\\
&=\xi_{k,m}\int_{-\Delta\theta_\mathrm{{max}}}^{\Delta\theta_\mathrm{{max}}}e^{j\alpha_{k,m}[\cos(\hat{\theta})\cdot\frac{1}{2}(\Delta\theta)^2-\sin(\hat{\theta})(\Delta\theta)]}p(\Delta\theta)d(\Delta\theta)\nonumber\\
&=\xi_{k,m}\int_{-\Delta\theta_\mathrm{{max}}}^{\Delta\theta_\mathrm{{max}}}\{\cos(\alpha_{k,m}\psi)+j\sin(\alpha_{k,m}\psi)\}p(\Delta\theta)d(\Delta\theta)\nonumber\\
&=\zeta_{k,m}+j\eta_{k,m}
\end{align}
where
\begin{align}
\xi_{k,m}=e^{-j\alpha_{k,m}\cos(\hat{\theta})},
\end{align}
and
\begin{align}
\psi=[\cos(\hat{\theta})\cdot\frac{1}{2}(\Delta\theta)^2-\sin(\hat{\theta})(\Delta\theta)].
\end{align}
Similarly, by utilizing the second-order Taylor expansion, $\cos(\alpha_{k,m}\psi)$ can be represented as
\begin{align}\label{Cos-Tay-Exp}
&\cos(\alpha_{k,m}\psi)=cos(\alpha_{k,m}[\cos(\hat{\theta})\cdot\frac{1}{2}(\Delta\theta)^2-\sin(\hat{\theta})(\Delta\theta)])\nonumber\\
&\approx1-\frac{1}{2}\alpha_{k,m}^2[\cos(\hat{\theta})\cdot\frac{1}{2}(\Delta\theta)^2-\sin(\hat{\theta})(\Delta\theta)]^2\nonumber\\
&=1-\frac{1}{2}\alpha_{k,m}^2\nonumber\\
&\cdot[\frac{1}{4}\cos(\hat{\theta})^2(\Delta\theta)^4+\sin(\hat{\theta})^2(\Delta\theta)^2-\cos(\hat{\theta})\sin(\hat{\theta})\cdot(\Delta\theta)^3]\nonumber\\
&=1-\frac{1}{8}\alpha_{k,m}^2\cos(\hat{\theta})^2(\Delta\theta)^4-\frac{1}{2}\alpha_{k,m}^2\sin(\hat{\theta})^2(\Delta\theta)^2\nonumber\\
&+\frac{1}{2}\alpha_{k,m}^2\cos(\hat{\theta})\sin(\hat{\theta})(\Delta\theta)^3
\end{align}
Since the last term of (\ref{Cos-Tay-Exp}) is an odd function of $\Delta\theta$, then
\begin{align}
\int_{-\Delta\theta_\mathrm{{max}}}^{\Delta\theta_\mathrm{{max}}}\frac{1}{2}\alpha_{k,m}^2\cos(\hat{\theta})\sin(\hat{\theta})(\Delta\theta)^3\cdot p(\Delta\theta)d(\Delta\theta)=0.
\end{align}
Now, let us define
\begin{align}
&\chi_1=\int_{-\Delta\theta_\mathrm{{max}}}^{\Delta\theta_\mathrm{{max}}}(\Delta\theta)^4\cdot p(\Delta\theta)d(\Delta\theta)\nonumber\\
&=\frac{2}{K_d\sqrt{2\pi \sigma^2}}\{-\sigma^2\cdot\Delta\theta_\mathrm{{max}}^3\ e^{-\frac{\Delta\theta_\mathrm{{max}}^2}{2\sigma^2}}-3\sigma^4\Delta\theta_\mathrm{{max}}e^{-\frac{\Delta\theta_\mathrm{{max}}^2}{2\sigma^2}}\nonumber\\
&+\frac{3\sqrt{2\pi}}{2}\sigma^5 erf(\frac{\Delta\theta_\mathrm{{max}}}{\sqrt{2}\sigma})\},
\end{align}
and
\begin{align}
&\chi_2=\int_{-\Delta\theta_\mathrm{{max}}}^{\Delta\theta_\mathrm{{max}}}(\Delta\theta)^2\cdot p(\Delta\theta)d(\Delta\theta)\nonumber\\
&=\frac{2}{K_d\sqrt{2\pi\sigma^2}}\{-\sigma^2\Delta\theta_\mathrm{{max}}\cdot e^{-\frac{\Delta\theta_\mathrm{{max}}^2}{2\sigma^2}}+\frac{\sqrt{2\pi}}{2}\sigma^3erf(\frac{\Delta\theta_\mathrm{{max}}}{\sqrt{2}\sigma})\}.
\end{align}
Then the real part $\zeta_{k,m}$ of (\ref{Ing-Analog}) can be expressed as
\begin{align}
\zeta_{k,m}=\xi_{k,m}(K_d-\frac{1}{8}\alpha_{k,m}^2\cos(\hat{\theta})^2\chi_1-\frac{1}{2}\alpha_{k,m}^2\sin(\hat{\theta})^2\chi_2)
\end{align}
In the same manner, we can have the $\eta_{k,m}$ as follows
\begin{align}
\eta_{k,m}=\frac{1}{2}\xi_{k,m}\cdot\alpha_{k,m}cos(\hat{\theta})\chi_2
\end{align}
Until now, we complete the derivation of $\hat{v}_{k,m}$. That is, $\hat{v}_{k,m}=\zeta_{k,m}+j\eta_{k,m}$. Due to the special structure of analog precoder, we only need the phase of $\hat{v}_{k,m}$. Therefore we can reformulated the analog   $v_{k,m}$ as
\begin{align}
v_{RAB, k,m}=\frac{1}{\sqrt{M}}\exp(j*\angle(\hat{v}_{k,m})).
\end{align}
which is what we need.

\ifCLASSOPTIONcaptionsoff
\newpage
\fi
\bibliographystyle{IEEEtran}
\bibliography{IEEEfull,REF}
\end{document}